\documentclass[a4paper, 11pt]{article}

\usepackage{jheppub} 
\usepackage{graphicx}
\usepackage{bm, array, slashed, braket, tensor, derivative} 
\usepackage{comment}
\usepackage{ulem}
\usepackage{hyperref}
\usepackage{xcolor}
\hypersetup{colorlinks=true} 
\renewcommand{\emph}[1]{\textit{#1}} 

\newcommand{\lambdabar}{{\mkern0.75mu\mathchar '26\mkern -9.75mu\lambda}} 
\newcommand{\Refl}{\mathrm{refl}} 
\newcommand{\Trans}{\mathrm{trans}} 
\newcommand{\Ret}{\mathrm{ret}} 

\begin{document} 

\title{Dynamically assisted Klein tunneling in the Furry picture} 
\author{Makoto Ochiai} 
\author{and Shunsuke Shibayama} 
\affiliation{Department of Physics, Waseda University, 169-8555 Tokyo, Japan} 
\emailAdd{ochiai36@akane.waseda.jp}
\emailAdd{sbym1346@fuji.waseda.jp} 
\date{April 14, 2025} 

\abstract{One-dimensional scattering of a wave packet of a relativistic fermion under a temporally oscillating electric field superimposed on a potential step is discussed by using the Furry-picture perturbation theory, where the oscillating electric field is treated as a perturbation. Reflection and transmission probabilities of the wave packet, which in its single-mode limit are consistent with those in the stationary scattering off the potential step alone, are investigated up to the second order. We show that even in the absence of the so-called Klein region, a positive-frequency incoming wave can penetrate the negative-frequency region below the potential step by emitting its energy to the oscillating electric field with a finite tunneling probability.} 


\maketitle

\section{\label{Sec1} Introduction}

One of the most fascinating features in quantum mechanics is the tunneling effect in potential scattering. In the case of stationary scattering under a time-independent potential barrier, a wave function satisfying the Schr\"{o}dinger equation can penetrate to the other side of the barrier even if the incident energy is lower than the potential height. This property stems from the fact that the wave function inside the potential barrier decays exponentially, but not to zero, for a finite potential width. It means, however, that if the potential barrier has an infinite width, such as a one-dimensional step potential, the tunneling does not occur, and the incident wave is totally reflected. In addition, as the potential barrier becomes higher, the exponential suppression of the wave function inside the barrier becomes stronger, and it becomes zero in the limit of infinite height, i.e., the rigid-wall limit. In relativistic quantum mechanics, however, the Dirac equation or the Klein--Gordon equation in such a potential can admit a solution that is nonvanishing in the limit of the infinite width and/or height~\cite{klein_1929_Z.Phys._ReflexionElektronenPotentialsprung}. This is because the so-called positive- and negative-energy regions (positive- and negative-frequency regions) are deformed by the potential, and tunneling can occur between these regions. It was pointed out soon after the proposal of the Dirac equation and is known as Klein tunneling~\cite{dombey.calogeracos_1999_Phys.Rep._SeventyYearsKlein, calogeracos.dombey_1999_Contemp.Phys._HistoryPhysicsKlein, itzykson.zuber_1980_book_QuantumFieldTheory, bjorken.drell_1964_book_RelativisticQuantumMechanics}. Although it seems to contradict our physical intuition, recent experiments using graphene materials have verified the effect~\cite{young.kim_2009_Nat.Phys._QuantumInterferenceKlein, stander.huard.ea_2009_Phys.Rev.Lett._EvidenceKleinTunneling}, which has attracted much attention in various fields~\cite{geim_2009_Science_GrapheneStatusProspects, geim.novoselov_2007_NatureMaterials_RiseGraphene, castroneto.guinea.ea_2009_Rev.Mod.Phys._ElectronicPropertiesGraphene, beenakker_2008_Rev.Mod.Phys._AndreevReflectionKlein, cooper.danjou.ea_2012_ISRNCond.Matt.Phys._ExperimentalReviewGraphene, abergel.apalkov.ea_2010_Adv.Phys._PropertiesGrapheneTheoretical}.

Klein tunneling has been considered a relevant phenomenon to the Schwinger effect~\cite{schwinger_1951_Phys.Rev._GaugeInvarianceVacuum} (see also \cite{fedotov.ilderton.ea_2023_Phys.Rep._AdvancesQEDIntense, ruffini.vereshchagin.ea_2010_Phys.Rep._ElectronPositronPairs, gelis.tanji_2016_Prog.Part.Nucl.Phys._SchwingerMechanismRevisited, schwartz_2013_book_QuantumFieldTheory, greiner.muller.ea_1985_book_QuantumElectrodynamicsStrong, itzykson.zuber_1980_book_QuantumFieldTheory, bjorken.drell_1964_book_RelativisticQuantumMechanics, fradkin.gitman.ea_1991_book_QuantumElectrodynamicsUnstable}). It is one of the non-perturbative effects in quantum field theory where the vacuum becomes unstable under an extremely strong electric field, from which particle--anti-particle pairs are spontaneously created. Its mechanism is roughly explained as follows: the electric field provides energy to virtual particle--anti-particle pairs in the vacuum to become real pairs. The number of particles produced per unit time and volume in a constant uniform electric field can be calculated exactly. The electric field along, say, the $z$-axis is given by a static scalar potential that is linear in the spatial coordinate as $A_0 \propto |\bm{E}| z$. We mention the work of Nikishov~\cite{nikishov_1969_JETP_PairProductionConstant, nikishov_1970_Nucl.Phys.B_BarrierScatteringField}, which reveals that the tunneling probability between the positive- and negative-frequency regions under the potential is closely related to the number of particles produced via the Schwinger effect. Hence, Klein tunneling is sometimes discussed in the framework of quantum field theory~\cite{nikishov_2004_Phys.Atom.Nucl._ScatteringPairProduction, hansen.ravndal_1981_Phys.Scr._KleinsParadoxIts, gavrilov.gitman_2016_Phys.Rev.D_ScatteringPairCreation, gavrilov.gitman_2016_Phys.Rev.D_QuantizationChargedFields, nakazato.ochiai_2022_Prog.Theor.Exp.Phys._UnstableVacuumFermion, krekora.su.ea_2004_Phys.Rev.Lett._KleinParadoxSpatial, krekora.su.ea_2005_Phys.Rev.A_KleinParadoxSpinresolved, cheng.su.ea_2010_Contemp.Phys._IntroductoryReviewQuantum, chervyakov.kleinert_2009_Phys.Rev.D_ExactPairProduction, chervyakov.kleinert_2018_Phys.Part.Nuclei_ElectronPositronPair}. 
%

It is known that the Schwinger effect exhibits various features depending on the configuration of electric fields. One of the most interesting phenomena discovered recently is the dynamically assisted Schwinger effect~\cite{schutzhold.gies.ea_2008_Phys.Rev.Lett._DynamicallyAssistedSchwinger}, in which the number of produced particles drastically increases in the presence of a weak and oscillating electric field in addition to a strong electric field. Since the ordinary Schwinger effect is exponentially suppressed with respect to the inverse of the electric field strength, i.e., $mc^2/(|e\bm{E}| \lambdabar)$, with a particle's mass $m$, charge $e$, and Compton wavelength $\lambdabar$, it is quite difficult to observe it with today's experimental techniques. Here, the denominator of the dimensionless quantity can be interpreted as the work done by the uniform electric field to separate a virtual pair of particle and anti-particle by the Compton wavelength. It is far less than the rest mass $mc^2$ at a normal electric field strength. The essence of the dynamically assisted Schwinger effect is that a temporally oscillating electric field provides energy as a wave, and the energy, together with that of the strong electric field, can exceed the rest mass to achieve pair production. Thus, the dynamically assisted Schwinger effect is regarded as a cooperative phenomenon of the non-perturbative effect by the strong electric field and the perturbative effect by the oscillating field. The total number or the momentum distribution of pairs created can vary greatly depending on the configuration of the strong and oscillating electric fields~\cite{schneider.schutzhold_2016_J.HighEnerg.Phys._DynamicallyAssistedSauterSchwinger, torgrimsson.schneider.ea_2017_J.HighEnerg.Phys._DynamicallyAssistedSauterSchwinger, torgrimsson.schneider.ea_2018_Phys.Rev.D_SauterSchwingerPairCreation, akal.egger.ea_2019_Phys.Rev.D_SimulatingDynamicallyAssisted, taya_2019_Phys.Rev.D_FranzKeldyshEffectStrongField, huang.taya_2019_Phys.Rev.D_SpinDependentDynamicallyAssisted, taya.fujimori.ea_2021_J.HighEnerg.Phys._ExactWKBAnalysis}, and theoretical studies are ongoing. It has also attracted the attention of experimental physicists, and several experiments using intense lasers are planned for the near future; see reviews~\cite{ruffini.vereshchagin.ea_2010_Phys.Rep._ElectronPositronPairs, fedotov.ilderton.ea_2023_Phys.Rep._AdvancesQEDIntense}.

Inspired by the dynamically assisted Schwinger effect, we consider the scattering problem in relativistic quantum mechanics. Specifically, we discuss how the scattering behavior, including Klein tunneling, changes when an oscillating electric field is added to a stationary potential. As mentioned above, Klein tunneling is relevant to pair production from the vacuum in the presence of external fields. Given that pair production from the vacuum changes significantly when an oscillating electric field is superimposed, we also expect the scattering behavior to be greatly affected. Note that similar attempts have been made before by several researchers~\cite{queisser.schutzhold_2019_Phys.Rev.C_DynamicallyAssistedNuclear, kohlfurst.queisser.ea_2021_Phys.Rev.Res._DynamicallyAssistedTunneling, ryndyk.kohlfurst.ea_2024_Phys.Rev.Res._DynamicallyAssistedTunneling}, including the discoverers of the dynamically assisted Schwinger effect. However, these studies have been limited so far to non-relativistic scattering problems. This paper could be positioned as one of the extensions of the previous studies to relativistic problems. Here, we adopt the Furry-picture perturbation theory~\cite{fradkin.gitman_1981_Fortschr.Phys._FurryPictureQuantum, torgrimsson.schneider.ea_2017_J.HighEnerg.Phys._DynamicallyAssistedSauterSchwinger, taya_2019_Phys.Rev.D_FranzKeldyshEffectStrongField}, in which the oscillating electric field is treated perturbatively and the stationary potential non-perturbatively. This is often used in the analysis of the dynamically assisted Schwinger effect to capture the cooperative effects of non-perturbative and perturbative effects mentioned above. Using this approach, we consider a tunneling effect that can occur only when both the stationary potential and the oscillating electric field are present, which we call dynamically assisted Klein tunneling.

This paper is organized as follows: after the introductory section, the main settings of the paper are given in section~\ref{Sec2}. Then, we give a wave packet formalism in section~\ref{Sec3} because the conventional framework of stationary scattering is not simply applicable under time-varying potentials. Here, the scattering of a wave packet without an oscillating electric field is discussed as a review of Klein tunneling. In the subsequent two sections (section~\ref{Sec4} and~\ref{Sec5}), we show analytical results of the first- and second-order perturbations, respectively, and compare them with numerical simulations. Section~\ref{Sec6} is devoted to the summary, conclusion, and future work. Before proceeding to the next section, we comment on the terminologies Klein tunneling and Klein paradox. The latter is explained as the literal contradiction that the reflection probability becomes greater than one. The difference between them originates from boundary conditions on the stationary scattering solutions of the Dirac equation~\cite{klein_1929_Z.Phys._ReflexionElektronenPotentialsprung, calogeracos.dombey_1999_Contemp.Phys._HistoryPhysicsKlein, dombey.calogeracos_1999_Phys.Rep._SeventyYearsKlein, itzykson.zuber_1980_book_QuantumFieldTheory, greiner.muller.ea_1985_book_QuantumElectrodynamicsStrong}, which is summarized in appendix~\ref{AppendA}. We also give the basic settings of the numerical simulations in section~\ref{Sec4} and~\ref{Sec5} in the appendix. Some formulae used in the wave packet formalism are shown in Appendix~\ref{AppendB}.

\section{\label{Sec2} Setup}

Throughout the paper, we adopt the natural units $\hbar = c = 1$. For simplicity, we restrict ourselves to the one-dimensional scattering, i.e., the spatial dependence of the potential is one-dimensional (e.g., $z$-dependent), and consider only the case of incident waves traveling parallel to the $z$-axis. Ignoring spatial components other than $z$, we shall write the space-time coordinates as $x = (t,z)$
\footnote{
More precisely, the $x$ and $y$ components of the wave function can be expressed as plane waves due to the translational symmetry, and we restrict ourselves to the momenta along the $x$ and $y$ axes to be zero. For this case, the spin along the $z$-direction is always conserved because the $z$-component of the Pauli matrix $\sigma_z$ commutes with the Dirac Hamiltonian.
}. Suppose that the step potential with height $V_0$ is given as a scalar potential:
\begin{align}
    V(z) = -eA_0 (z) = V_0 \, \theta (z). \label{eq:(sec2)V(z)} 
\end{align}
Here, $e > 0$ is the magnitude of the electric charge, and $\theta (z)$ is the Heaviside step function. The potential gives an electric field localized at $z = 0$, i.e., $-V_0 \, \delta (z)$. For relativistic fermions of mass $m$, the Dirac equation under the step potential \eqref{eq:(sec2)V(z)} can be written by using the Dirac gamma matrices $\gamma^0, \gamma^3$ as 
\begin{align}
    [i\gamma^0 (\partial_t - ieA_0 (z)) + i\gamma^3 \partial_z - m] \psi (x) = 0. \label{eq:(sec2)DiracEq_step} 
\end{align}
When we explore a solution with an energy eigenvalue $E$ and spin $s$ as 
\begin{align}
    \psi (x) = e^{-iEt} \psi_s^{(E)} (z), 
\end{align}
we can separate the variables $t$ and $z$ to obtain the stationary Dirac equation 
\begin{align}
    [-i\gamma^0 \gamma^3 \partial_z + m\gamma^0 + V(z)] \psi_s^{(E)} (z) = E\psi_s^{(E)} (z). \label{eq:(sec2)StationaryDiracEq} 
\end{align}
Equation \eqref{eq:(sec2)StationaryDiracEq} can be exactly solved as in the first course of quantum mechanics. That is, there are plane-wave solutions on the left ($z < 0$) and right ($z > 0$) of the step, respectively, and the solutions should be connected so that the eigenfunctions are continuous at $z = 0$. In this case, there are positive- or negative-frequency scattering states belonging to the energy eigenvalues $|E| > m$ or $|E - V_0| > m$. In particular, when the potential height is greater than twice the fermion mass ($V_0 > 2m$: overcritical), the so-called Klein region $m < E < V_0 - m$ arises where positive-frequency solutions on the left of the step are connected to negative-frequency solutions on the right. Therefore, the eigenstates show oscillatory behavior on both sides of the step, leading to a nonvanishing transmission probability. One of the surprising features of Klein tunneling is that the transmission probability is nonvanishing even in the rigid-wall limit ($V_0 \to \infty$) of the potential, which is never seen in non-relativistic quantum mechanics; see the next section. Unlike the Klein paradox, note that the reflection probability does not exceed one.

In addition to the step potential \eqref{eq:(sec2)V(z)}, we consider a space-time dependent vector potential:
\begin{align}
    A_3 (x) = \frac{\mathcal{E}_z}{\omega} \sin (\omega t) e^{-(z/l)^2}. \label{eq:(sec2)A3(x)} 
\end{align}
$A_3 (x)$ is localized at $z = 0$ around the electric field with a Gaussian width $l$, giving an oscillating electric field with frequency $\omega$ and maximum field strength $\mathcal{E}_z$. The time-dependence of $A_3 (x)$ is intended to promote the tunneling process, as the energy of an oscillating electric field promotes the Schwinger pair production. In addition, the spatial localization factor is assumed so that a wave packet, which is introduced in the next section, is affected by the vector potential only in a finite spatial region $|z| \sim l$.

We will work on the scattering problem under the potentials \eqref{eq:(sec2)V(z)} and \eqref{eq:(sec2)A3(x)}, but since the vector potential $A_3(x)$ is time-dependent, the framework of stationary scattering cannot be used. Therefore, we consider a wave packet approach instead. In what follows, we will distinguish between a scattering wave function and a wave packet by denoting them as $\psi$ and $\Psi$.

Under the two external fields, the Dirac equation is
\begin{align}
    [i\gamma^0 (\partial_t - ieA_0 (z)) + i\gamma^3 \partial_z - m] \Psi (x) = -e\gamma^3 A_3 (x) \Psi (x). \label{eq:(sec2)DiracEq_full} 
\end{align}
We have to solve the Dirac equation~\eqref{eq:(sec2)DiracEq_full} to compute the reflection and transmission probabilities. To do this, we will consider wave packet scattering in the next section and then define the reflection and transmission probabilities. When solving the full Dirac equation~\eqref{eq:(sec2)DiracEq_full}, the step potential~\eqref{eq:(sec2)V(z)} is treated non-perturbatively and the oscillating field~\eqref{eq:(sec2)A3(x)} is incorporated perturbatively (Furry-picture perturbation \cite{fradkin.gitman_1981_Fortschr.Phys._FurryPictureQuantum, torgrimsson.schneider.ea_2017_J.HighEnerg.Phys._DynamicallyAssistedSauterSchwinger, taya_2019_Phys.Rev.D_FranzKeldyshEffectStrongField}).
More precisely, we choose the dimensionless perturbation parameter as 
\begin{align}
    \lambda = \frac{e\mathcal{E}_z}{2 m \omega} 
\end{align}
and calculate the wave function in power series of $\lambda \ll 1$, where the perturbation characterized by $e\mathcal{E}_z/\omega$ is much smaller than the mass; see section~\ref{Sec4} for details.

\section{\label{Sec3}Scattering of a wave packet}

There are two basic approaches to scattering problems in quantum theory. One is to naively consider the scattering process in which an incident particle hits a target particle as a dynamical process. It is non-stationary scattering, where we consider wave packets; see the formal theory of wave packet scattering~\cite{sunakawa_1955_Prog.Theor.Phys._FormalTheoryScattering, namiki.iino_1958_Prog.Theor.Phys.Suppl._NewMathematicalFormulation} and numerical simulations of the wave packet scattering under the Klein step~\cite{nitta.kudo.ea_1999_Am.J.Phys._MotionWavePacket, leo.rotelli_2006_Phys.Rev.A_BarrierParadoxKlein, krekora.su.ea_2004_Phys.Rev.Lett._KleinParadoxSpatial, krekora.su.ea_2005_Phys.Rev.A_KleinParadoxSpinresolved, cheng.su.ea_2010_Contemp.Phys._IntroductoryReviewQuantum}. The other is to focus on the stationary flow of incident and scattered particles that are created by an incident flow of particles. The latter approach is convenient for stationary scattering problems and is often used. Since there is now a time-dependent vector potential, the scattered particles do not create a stationary flow in a naive way (however, since eq.~\eqref{eq:(sec2)A3(x)} is periodic in time, it may be possible to treat it by using the Floquet theory; for example, a recent study~\cite{ryndyk.kohlfurst.ea_2024_Phys.Rev.Res._DynamicallyAssistedTunneling} discusses non-relativistic scattering by using this method). In this section, we define the reflection and transmission probabilities on the basis of the scattering of wave packets. Then, as an example, we calculate those probabilities under the step potential alone and see that the results are consistent with those in stationary scattering. We also discuss the Klein tunneling of the wave packet by an overcritical potential step ($V_0 > 2m$).

\begin{figure}[htbp]
    \centering
    \includegraphics[width=0.6\linewidth]{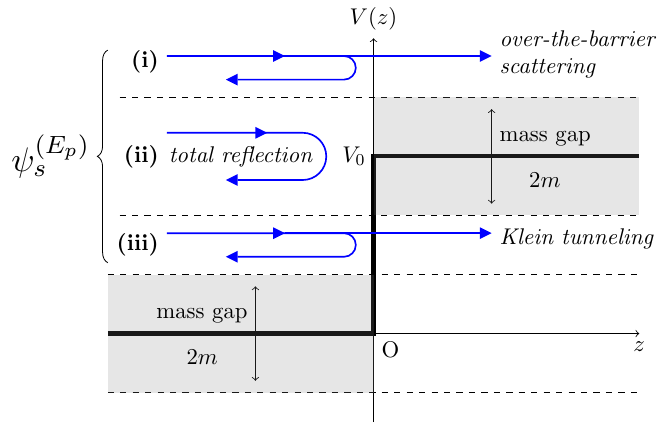} 
    \caption{\label{fig:psi_s_Ep}Scattering behavior of the left-incident and 
 positive-frequency solution $\psi_s^{(E_p)}$ off the overcritical potential step. Gray-shaded regions represent the forbidden region or the mass gap. Blue arrows for (i), (ii), or (iii) show the directions of the incident, reflected, and transmitted waves for each.}
\end{figure}

Suppose a positive-frequency wave packet is incident from the far left ($z = -\infty$). The wave packet is realized as a suitable superposition of the so-called scattering wave functions. The scattering wave function is a solution of the stationary Dirac equation \eqref{eq:(sec2)StationaryDiracEq} under the step potential alone, which is characterized by positive or negative frequencies and boundary conditions, i.e., whether the incident plane wave comes from the far left or the far right $z = \pm \infty$. We denote the positive-frequency left-incident solution as $\psi^{(E_p)}_s (x)$, where $E_p = \sqrt{p^2 + m^2}$ is the on-shell energy eigenvalues and $s$ is the spin. $p \, (> 0)$ denotes the momentum of the incident plane wave on the left side of the step. $\psi^{(E_p)}_s$ shows different behavior depending on the following three energy regions (see also figure~\ref{fig:psi_s_Ep}): 
\begin{itemize}
    \item[(i)] Over-the-barrier scattering for $E_p \geq V_0 + m$ \\ 
    When the energy eigenvalues are higher than the mass gap on the right side of the step, the incident wave is partially reflected at $z = 0$, and its remainder transmits on the right side of the step, as in non-relativistic quantum mechanics. Both reflected and transmitted waves have positive frequencies. 
    \item[(ii)] Total reflection for $V_0 - m \leq E_p < V_0 + m$ \\ 
    As the energy eigenvalues are lowered, total reflection occurs as in ordinary quantum mechanics. Due to the mass gap on the right side of the step, there is no transmitted wave, and the solution in the region $z > 0$ is exponentially damped. Note that for the subcritical step ($V_0 < 2m$), the lower bound of the energy eigenvalues is replaced by $m$ instead of $V_0 - m$.
    \item[(iii)] Klein tunneling for $m \leq E_p < V_0 - m$ \\ 
    When the energy eigenvalues are in the Klein region, the incident and reflected waves are above the mass gap on the left, while the transmitted wave is below the mass gap on the right side. In other words, the positive- and negative-frequency modes are allowed, and they are connected at $z = 0$. Specifically,
    \begin{equation}\begin{split} 
        \psi_s^{(E_p)} (z) = \frac{1}{\sqrt{2\pi}} \sqrt{\frac{m}{E_p}} &\bigl\{ \theta (-z) \bigl[ u(p, s) e^{ipz} + R_\psi (p) u(-p, s) e^{-ipz} \bigr] \\ 
&+ \theta (z) T_\psi (p) v(q, s) e^{iqz} \bigr\}, \label{eq:(sec3)psi_KleinTunneling}
    \end{split}\end{equation}
    where $u$ and $v$ are the positive- and negative-frequency four-component Dirac spinors, $R_\psi(p)$ and $T_\psi(p)$ are the reflection and transmission coefficients. $q \, (> 0)$ is the momentum of the transmitted wave, which is determined by the energy conservation $E= E_p = V_0 - E_q$, where $E_q = \sqrt{q^2 + m^2}$. The spin $s$ is the label of the two-component spinor $\bm{\xi} (s)$ contained in the Dirac spinors $u$ and $v$, which satisfies $\sigma_z \bm{\xi} (s) = s \bm{\xi} (s)$. Since the $z$-component of the current generated by those scattering wave functions $j_z = \bar{\psi} \gamma^3 \psi$ should be continuous at $z = 0$, the so-called probability conservation relation holds: 
    \begin{align}
        |R_\psi (p)|^2 + \frac{q}{p} |T_\psi (p)|^2 = 1. \label{eq:(sec3)prob_conserbation_psi}
    \end{align}
    The first and second terms on the left-hand side correspond to the reflection and transmission probabilities, and particularly, the transmission probability becomes nonvanishing in the rigid-wall limit. It should be mentioned that the phase velocity of the transmitted wave is negative, while the group velocity and current $j_z$ are positive. On the other hand, the so-called Klein paradox discusses solutions with the positive phase velocity of the transmitted wave; see the appendix for details. Note also that the left-incident scattering wave functions are normalized by using the Dirac inner product as 
    \begin{align}
        \int_{-\infty}^\infty \odif{z} \psi_s^{(E_p) \dagger} (z) \psi_{s'}^{(E_{p'})} (z) = \delta (p - p') \delta_{s, s'}. 
    \end{align}
\end{itemize}
Note that all the scattering states, including the negative-frequency or the right-incident states, are summarized in ref.~\cite{ochiai.nakazato_2018_J.Phys.Commun._CompletenessScatteringStates}, which forms a complete orthonormal set.

We construct a wave packet by integrating these positive-frequency left-incident scattering wave functions with a weight $g(p)$ over the incident momentum $p$, i.e., 
\begin{align}
    \Psi^{(0)} (x) = \int_0^\infty \odif{p} \, g(p) \psi_s^{(E_p)} (x). \label{eq:(sec3)Psi0} 
\end{align}
The superscript ``(0)" on the left-hand side indicates the zeroth order of perturbation, or without the oscillating electric field \eqref{eq:(sec2)A3(x)}. Here, the spin $s$ is fixed for simplicity. For the same reason, we do not include right-incident and negative-frequency modes in the wave packet. The wave packet $\Psi^{(0)}$ satisfies the continuity relation in terms of the probability density and current density: 
\begin{align}
    \partial_t \rho^{(0)} + \partial_z J_z^{(0)} = 0, \label{eq:(sec3)current_continuity_Psi0} 
\end{align}
where
\begin{align}
    \rho^{(0)} (x) &= \Psi^{(0) \dagger} (x) \Psi^{(0)} (x), \label{eq:(sec3)rho0} \\ 
    J_z^{(0)} (x) &= \bar{\Psi}^{(0)} (x) \gamma^3 \Psi^{(0)} (x). \label{eq:(sec3)Jz0} 
\end{align}
This is one of the properties that follow from the Dirac equation under the step potential \eqref{eq:(sec2)DiracEq_step}. The probability density of the wave packet is always non-negative, and the integration over the whole space is time-independent. Given the normalization for the probability density 
\begin{align}
    \int_{-\infty}^\infty \odif{z} \rho^{(0)} (x) = 1, \label{eq:(sec3)normalization_rho(0)} 
\end{align}
we obtain the normalization condition for the weight function by substituting the expansion \eqref{eq:(sec3)Psi0} into the above equation and using the orthonormality of $\psi_s^{(E_p)}$, as 
\begin{align}
    \int_0^\infty \odif{p} |g(p)|^2 = 1. 
\end{align}
We take the weight function as a Gaussian form centered at momentum $\bar{p}$ and deviation $\sigma$, 
\begin{align}
    g(p) = N\exp \Bigl[ -\frac{(p - \bar{p})^2}{2\sigma^2} \Bigr] 
\end{align}
with the normalization factor 
\begin{align}
    N = \Bigl( \frac{\sqrt{\pi}}{2} \sigma \bigl[ 1 + \mathrm{erf} (\bar{p}/\sigma) \bigr] \Bigr)^{-1/2}, 
\end{align}
where $\mathrm{erf}$ denotes the error function. The momentum deviation is supposed to be sufficiently small compared to the central momentum. In particular, in the limit $\sigma \to 0$, $|g(p)|^2$ converges to a delta function $\delta (p - \bar{p})$, i.e., the wave packet becomes a single mode, spreading over the entire position space. Note that for the full Dirac equation \eqref{eq:(sec2)DiracEq_full}, we can derive the continuity relation in which the superscripts ``(0)" of eqs.~\eqref{eq:(sec3)rho0} and \eqref{eq:(sec3)Jz0} are removed: 
\begin{align} 
  \partial_t \rho + \partial_z J_z = 0. \label{eq:(sec3)current_continuity_Psi} 
\end{align} 
Since the oscillating electric field is localized around $z = 0$ of the step, it initially does not affect the wave packet $\Psi$, i.e., the probability density at the remote past satisfies $\rho (z, -\infty) = \rho^{(0)} (z, -\infty)$. Hence, the normalization condition for the probability density \eqref{eq:(sec3)normalization_rho(0)} implies 
\begin{align}
    \int_{-\infty}^\infty \odif{z} \rho (x) = 1. \label{eq:(sec3)normalization_rho} 
\end{align}

Now, we consider the reflection and transmission probabilities for the wave packet scattering. Let $z_L < 0$ and $z_R > 0$ be positions on $z$-axis which are sufficiently far from the step at $z = 0$, and integrate the continuity equation \eqref{eq:(sec3)current_continuity_Psi} over an interval $[z_L, z_R]$, i.e., 
\begin{align} 
  \odv{}{t} \int_{z_L}^{z_R} \odif{z} \rho (z, t) + J_z (z_R, t) - J_z (z_L, t) = 0. \label{eq:(sec3)current_continuity_integrated} 
\end{align} 
Suppose that we observe the incident and scattered currents of the wave packet by the detectors placed at $z = z_L$ and $z_R$ according to the following steps: 
\begin{itemize} 
  \item[1.] We turn on the detectors at the remote past $t = -\infty$. 
  \item[2.] The wave packet is injected from the left infinity $z = -\infty$. 
  \item[3.] We turn off the detectors at a finite time $t_0$ (which means that the detection of the initial wave packet is stopped). 
  \item[4.] At a later time $t_1 \, (> t_0)$, we turn on the detectors again to detect the reflected and transmitted wave packets until the remote future $t = \infty$. 
\end{itemize} 
In the third step, the total amount of $J_z$ observed by the detector on the left of the step is expressed as
\begin{equation}\begin{split}
    \int_{-\infty}^{t_0} \odif{t} J_z (z_L, t) &= \int_{-\infty}^{t_0} \odif{t} J_z (z_R, t_0) + \int_{z_L}^{z_R} \odif{z} \rho (z, t_0) = \int_{z_L}^\infty \odif{z} \rho (z, t_0), \label{eq:(sec3)t-integral_incident_current} 
\end{split}\end{equation}
where $\rho (z, -\infty) = 0$ has been used in the first equality because the wave packet at the remote past should be far left from $z_L$. It has also been used in the second equality that the second expression is $z_R$-independent and can be evaluated particularly in the limit $z_R \to \infty$, in which the first term should vanish because only the detector at $z_L$ can capture the incident current. Now, the incident wave packet cannot be fully detected because the detector is switched off at the finite time $t_0$. Furthermore, the incident and reflected wave packets give rise to interference, which will be included in the left-hand side of eq.~\eqref{eq:(sec3)t-integral_incident_current}. However, if we take the limit $z_L \to -\infty$ with fixing the switch-off time $t_0$, it is expected that only the incident wave packet can be fully detected and the interference effect can vanish, i.e., $\lim_{z_L \to -\infty} \int_{-\infty}^{t_0} \odif{t} J_z (z_L, t)$ is expected to be interpreted as the total amount of the incident wave packet. In the same way, the total amount of $J_z$ observed at $z_L$ and $z_R$ in the forth step can be written as
\begin{align}
    \int_{t_1}^\infty \odif{t} \bigl( J_z (z_R, t) - J_z (z_L, t) \bigr) &= \int_{z_L}^{z_R} \odif{z} \rho (z, t_1), \label{eq:(sec3)t-integral_scattered_current} 
\end{align}
where $\rho (z, \infty) = 0$ has been used. It is expected that in the limit $z_L \to -\infty$ and $z_R \to \infty$ with $t_1$ fixed, the left-hand side can be interpreted as the total amount of the the scattered wave packets. Observe that the right-hand sides of eqs.~\eqref{eq:(sec3)t-integral_incident_current} and \eqref{eq:(sec3)t-integral_scattered_current} in the limit $z_L \to -\infty$ and $z_R \to \infty$ coincide with each other, independently on the detection switch-on/off times $t_0$ and $t_1$. It thus implies that the total amount of the incident and scattered currents observed at the spatial infinity, which we are assuming, is conserved: 
\begin{equation}\begin{split}
    &\lim_{z_L \to -\infty} \int_{-\infty}^{t_0} \odif{t} J_z (z_L, t) = \lim_{\substack{z_L \to -\infty \\ z_R \to \infty}} \int_{t_1}^\infty \odif{t} \bigl( J_z (z_R, t) - J_z (z_L, t) \bigr). \label{eq:(sec3)probability_conservation} 
\end{split}\end{equation}
The above discussion holds regardless of the presence or absence of the oscillating electric field. Of course, the existence of such limits is a nontrivial problem. However, we will show that the computation of the above quantities brings about consistent results with those of conventional stationary scattering. Now, we define the reflection probability of the wave packet 
\begin{align}
    P_\Refl = \Biggl| \lim_{z_L \to -\infty} \frac{\int_{t_1}^\infty \odif{t} J_z (z_L, t)}{\int_{-\infty}^{t_0} \odif{t} J_z (z_L, t)} \Biggr|, \label{eq:(sec3)P_refl_def} 
\end{align}
as well as the transmission probability of the wave packet 
\begin{align}
    P_\Trans = \Biggl| \lim_{\substack{z_L \to -\infty \\ z_R \to \infty}} \frac{\int_{t_1}^\infty \odif{t} J_z (z_R, t)}{\int_{-\infty}^{t_0} \odif{t} J_z (z_L, t)} \Biggr|. \label{eq:(sec3)P_trans_def} 
\end{align}
Equation~\eqref{eq:(sec3)probability_conservation} ensures that the sum of the probabilities is normalized. Since the probability density $\rho$ is normalized as in eq.~\eqref{eq:(sec3)normalization_rho}, the denominator on the right-hand side of eqs.~\eqref{eq:(sec3)P_refl_def} and \eqref{eq:(sec3)P_trans_def} is reduced to one.

Let us compute in practice the reflection and transmission probabilities in the absence of the oscillating electric field: $P_\Refl^{(0)}$ and $P_\Trans^{(0)}$. First, we perform the integral of $J_z^{(0)} (z_L, t)$ over time from $t = t_1$ to infinity. Substituting the specific form of the mode function $\psi_s^{(E_p)}$ into the wave packet $\Psi^{(0)}$, we obtain the following expression: 
\begin{equation}\begin{split}
    &\int_{t_1}^\infty \odif{t} J_z^{(0)} (z_L, t) \\ 
    &= \int_0^\infty \frac{\odif{p} \odif{p'}}{2\pi} \frac{m}{\sqrt{E_p E_{p'}}} g^* (p) g(p') \frac{i}{E_p - E_{p'} + i0_+} e^{i(E_p - E_{p'}) t_1} \\ 
    &\quad \times \Bigl\{ \bar{u} (p, s) \gamma^3 u(p', s) e^{-i(p - p') z_L} + R_\psi^* (p) R_\psi (p') \bar{u} (-p, s) \gamma^3 u(-p', s) e^{i(p - p') z_L} \\ 
    &\qquad + R_\psi (p') \bar{u} (p, s) \gamma^3 u(-p',s) e^{-i(p + p') z_L} + R_\psi^* (p) \bar{u} (-p, s) \gamma^3 u(p', s) e^{i(p + p') z_L} \Bigr\}. \label{eq:(sec3)t-integral_JzL(0)_calc} 
\end{split}\end{equation}
The distribution $i/(E_p - E_{p'} + i0_+)$, where $0_+$ is a positive infinitesimal, appears by integrating the time-dependent exponential factor of the mode function over time from $t = t_1$ to infinity 
\footnote{
This calculation is conducted by adding the convergence factor $\exp [-0_+t]$ to the integrand, which implicitly assumes the physical context where a detector placed on the left of the step will no longer capture the wave packet in the infinite future. In other words, the scattered wave packet simply passes through the detector and travels off into spatial infinity $z = \pm \infty$; it will not be reflected off the boundary and captured again by the detector. This is the boundary condition for spatial infinity.
} . The first two terms in the parentheses are the diagonal terms that arise from the incident and reflected wave packets, and the remaining two terms are the cross terms causing interference. It implies that if the detector is placed at the finite position $z_L$, not only the contribution of the reflected wave packet but also the contribution of the incident wave packet and the interference effect will be captured. If we take the far-left limit $z_L \to -\infty$, however, the interference terms, which contain indefinitely oscillating factors $\exp [\pm i(p + p') z_L]$, vanish due to the Riemann--Lebesgue lemma. In addition, we can utilize the following formulae for the distribution of energy in the far-left limit, i.e, 
\begin{align}
    \lim_{z_L \to -\infty} \frac{i}{E_p - E_{p'} + i0_+} e^{-i(p - p') z_L} &= 0, \label{eq:(sec3)limit_formula_1} \\ 
    \lim_{z_L \to -\infty} \frac{i}{E_p - E_{p'} + i0_+} e^{i(p - p') z_L} &= 2\pi \delta (E_p - E_{p'}). \label{eq:(sec3)limit_formula_2}
\end{align}
(The derivation of the formulae is given in appendix~\ref{AppendB}.) Hence, only the contribution of the reflected wave packet survives. As a result, the numerator on the right-hand side of eq.~\eqref{eq:(sec3)P_refl_def} in the far-left limit becomes 
\begin{align}
    \lim_{z_L \to -\infty} \int_{t_1}^\infty J_z^{(0)} (z_L, t) = -\int_0^\infty \odif{p} |g(p)|^2 |R_\psi (p)|^2, 
\end{align}
where $\bar{u} (-p, s) \gamma^3 u(-p, s) = -p/m$ has been used. It should be remarked that the dependence on the detection switch-on time $t_1$ included in the time integral of the current disappears in the far-left limit. This implies that if the detector is placed infinitely far from the step at $z = 0$, the reflected wave packet can be detected regardless of the detection switch-on time. It can also be checked that the denominator in the right-hand side of eqs.~\eqref{eq:(sec3)P_refl_def} and \eqref{eq:(sec3)P_trans_def} converges to one by a similar calculation, where only the contribution of the incident wave packet survives. Therefore, the reflection probability of the wave packet under the potential step is 
\begin{align}
    P_\Refl^{(0)} = \int_0^\infty \odif{p} |g(p)|^2 |R_\psi (p)|^2. 
\end{align}
This is consistent with the reflection probability in stationary scattering in the plane-wave limit $|g(p)|^2 \to \delta (p - \bar{p})$.

For the transmission probability, we evaluate the far-right limit $z_R \to \infty$ of the time integral of $J_z^{(0)} (z_R, t)$. For the energy range $E_p \in \text{(i)}$ or (iii), it is convenient to use the formula in which $p$ is replaced with $q$ and $z_L$ with $z_R$ in eq.~\eqref{eq:(sec3)limit_formula_2}, i.e., 
\begin{align}
    \lim_{z_R \to \infty} \frac{i}{E_q - E_{q'} + i0_+} e^{-i(q - q') z_R} = 2\pi \delta (E_q - E_{q'}). 
\end{align}
Taking the energy conservation relation $E_p = V_0 \pm E_q$ into account, we obtain 
\begin{align}
    \lim_{z_R \to \infty} \int_{t_1}^\infty \odif{t} J_z^{(0)} (z_R, t) = \int_{|E_p - V_0| > m} \odif{p} |g(p)|^2 \frac{q}{p} |T_\psi (p)|^2. 
\end{align}
The result does not include the contribution of the energy range $E_p \in \text{(ii)}$ because the scattering wave function belonging to the range decays exponentially in $z$. Thus, the transmission probability $P_\Trans^{(0)}$ of the wave packet is 
\begin{align}
    P_\Trans^{(0)} = \int_{|E_p - V_0| > m} \odif{p} |g(p)|^2 \frac{q}{p} |T_\psi (p)|^2. 
\end{align}
This is also consistent with the results of stationary scattering. It is now obvious that the sum of the reflection and transmission probabilities of the wave packet becomes unity: 
\begin{align}
    P_\Refl^{(0)} + P_\Trans^{(0)} = 1. 
\end{align}
This is guaranteed by the probability conservation relation \eqref{eq:(sec3)prob_conserbation_psi} for the stationary scattering wave function.

\section{\label{Sec4}First-order perturbation in the Furry picture}

From now on, we will discuss the wave-packet scattering in the presence of not only the potential step but also the oscillating field~\eqref{eq:(sec2)A3(x)}. We will investigate the problem by making use of the Furry-picture perturbation theory, where the right-hand side of the evolution equation~\eqref{eq:(sec2)DiracEq_full} is treated as a perturbation. The wave packet $\Psi (x)$ is expressed up to the first-order perturbation as 
\begin{align}
    \Psi (x) = \Psi^{(0)} (x) + \Psi^{(1)} (x) + \mathcal{O} (\lambda^2). 
\end{align}
The first term of the right-hand side is the same as that in the previous section. The next term is
\begin{align}
    \Psi^{(1)} (x) = -e\int \odif[order=2]{x'} S_\mathrm{ret} (x, x') \gamma^3 A_3 (x') \Psi^{(0)} (x'), \label{eq:(sec4)Psi(1)} 
\end{align}
where the space-time integration is performed over $x' = (z', t')$, and $S_\Ret (x, x')$ is the retarded Green function, i.e., the solution of the equation where the interaction term on the right-hand side of eq.~\eqref{eq:(sec2)DiracEq_full} is replaced with the delta-function $\delta^2 (x - x')$. In particular, the retarded Green function can be expressed by using the complete set of solutions of the homogeneous equation~\eqref{eq:(sec2)DiracEq_step} as 
\begin{equation}\begin{split} 
    S_\Ret (x, x') &= -i\theta (t - t') \sum_{s, \epsilon = \pm} \biggl[ \int_0^\infty \odif{p} \, \psi_s^{(\epsilon E_p)} (x) \bar{\psi}_s^{(\epsilon E_p)} (x') + \int_0^\infty \odif{q} \phi_s^{(V_0 + \epsilon E_q)} (x) \bar{\phi}_s^{(V_0 + \epsilon E_q)} (x') \biggr]. \label{eq:(sec4)S_ret}
\end{split}\end{equation} 
Here $\psi_s^{(\epsilon E_p)}$ and $\phi_s^{(V_0 + \epsilon E_q)}$ are the left-incident and right-incident scattering wave functions, with $p$ and $q$ representing momenta on the left and right of the step, respectively. These states have positive- or negative-frequency modes, which are characterized by $\epsilon = +$ or $-$, and the scattering wave functions are classified into five types (under the overcritical step) depending on their energy eigenvalues: 
\begin{itemize}
    \item[(i)] For $E \geq V_0 + m$, the left- and right-incident solutions $\psi_s^{(E_p)}$ and $\phi_s^{(V_0 + E_q)}$, which describe the over-the-barrier scattering of positive-frequency waves, have the same energy ($E = E_p = V_0 + E_q$). 
    \item[(ii)] For $V_0 - m \leq E < V_0 + m$, a unique solution $\psi_s^{(E_p)}$ exists and describes the total reflection. 
    \item[(iii)] For $m \leq E < V_0 - m$, not only the left-incident solution $\psi_s^{(E_p)}$ \eqref{eq:(sec3)psi_KleinTunneling} but also the right-incident one $\phi_s^{(V_0 - E_q)}$ are present ($E = E_p = V_0 - E_q$), which bring about Klein tunneling. 
    \item[(iv)] For $-m \leq E < m$, only the right-incident scattering state $\phi_s^{(V_0 - E_q)}$ exists, where a negative-frequency wave is totally reflected on the right of the step. 
    \item[(v)] For $E < -m$, the eigenstates $\psi_s^{(-E_p)}$ and $\phi_s^{(V_0 - E_q)}$, which describe the over-the-barrier scattering of negative-frequency waves, have the same energy ($E = -E_p = V_0 - E_q$). 
\end{itemize}
As mentioned in the previous section, the Klein region (iii) does not appear under the subcritical case. Accordingly, the lower bound of the region (ii) is replaced with $m$, and the upper bound of the region (iv) with $V_0 - m$. Thus, the motion of the wave packet is perturbed by the oscillating field via the retarded Green function~\eqref{eq:(sec4)S_ret}. The normalization factor for $\Psi$ is the same as that for $\Psi^{(0)}$ because the retarded Green function vanishes in the remote past, which means $\Psi \to \Psi^{(0)}$ in the limit. It should be remarked that the Green function carries not only perturbative effects of the oscillating field but also non-perturbative effects that stem from the scattering wave functions under the potential step.

Expanding the current $J_z$ of the wave packet up to the first order reads 
\begin{align}
    J_z (x) = J_z^{(0)} (x) + J_z^{(1)} (x) + \mathcal{O} (\lambda^2), 
\end{align}
where $J_z^{(1)}$ is a current composed of the zeroth- and first-order wave packets 
\begin{align}
    J_z^{(1)} (x) = 2\Re \bigl\{ \bar{\Psi}^{(0)} (x) \gamma^3 \Psi^{(1)} (x) \bigr\}. 
\end{align}
In the same way as in section~\ref{Sec3}, we will evaluate the time integral of $J_z^{(1)}$ in the far-left and far-right limits to compute the reflection and transmission probabilities of the wave packet. Substitution of eqs.~\eqref{eq:(sec2)A3(x)}, \eqref{eq:(sec3)Psi0}, \eqref{eq:(sec4)Psi(1)}, and \eqref{eq:(sec4)S_ret} into the time integral of $\bar{\Psi}^{(0)} \gamma^3 \Psi^{(1)}$ at $z = z_L$ gives 
\begin{equation}\begin{split}
    &\int_{t_1}^\infty \odif{t} \bar{\Psi}^{(0)} (z_L, t) \gamma^3 \Psi^{(1)} (z_L, t) \\ 
    &= \lambda \int_0^\infty \odif{k} \odif{p} g^* (k) g(p) \sum_{s', \epsilon} \\ 
    &\quad \times \biggl[ \int_0^\infty \odif{p'} \bar{\psi}_s^{(E_k)} (z_L) \gamma^3 \psi_{s'}^{(\epsilon E_{p'})} (z_L) \braket{\psi_{s'}^{(\epsilon E_{p'})}, \psi_s^{(E_p)}} f_{\psi, \epsilon} (k, p', p; \omega; t_1) \\ 
    &\qquad + \int_0^\infty \odif{q'} \bar{\psi}_s^{(E_k)} (z_L) \gamma^3 \phi_{s'}^{(V_0 + \epsilon E_{q'})} (z_L) \braket{\phi_{s'}^{(V_0 + \epsilon E_{q'})}, \psi_s^{(E_p)}} f_{\phi, \epsilon} (k, q', p; \omega; t_1) \biggr]. \label{eq:(sec4)t-integral_Psi(0)bar_gamma_Psi(1)}
\end{split}\end{equation} 
Here $k$ denotes the momentum of the zeroth-order wave packet $\bar{\Psi}^{(0)} (z_L, t)$, while $p$ the momentum of $\Psi^{(0)} (x')$ included in the first-order wave packet $\Psi^{(1)} (z_L, t)$. The momenta $p', q'$ and spin $s'$ stem from the Green function $S_\Ret$. We have introduced the angle-bracket form $\braket{\psi_1, \psi_2}$ as a shorthand notation of an overlap integral of two scattering wave functions $\psi_1, \psi_2$ and the oscillating gauge field: for instance, the angle-bracket form in the third line represents 
\begin{align}
    \braket{\psi_{s'}^{(\epsilon E_{p'})}, \psi_s^{(E_p)}} = m \int_{-\infty}^\infty \odif{z} \bar{\psi}_{s'}^{(\epsilon E_{p'})} (z) \gamma^3 \psi_s^{(E_p)} (z) e^{-(z/l)^2}. 
\end{align}
This factor characterizes the strength of the coupling between the gauge field and two fermion external lines, and these external lines are affected (or referred to as "dressed") by the stationary potential non-perturbatively. We have also expressed the time-dependent parts as $f_{\psi, \epsilon}$ and $f_{\phi, \epsilon}$, i.e., 
\begin{equation}\begin{split} 
    &f_{\psi, \epsilon} (k, p', p; \omega; t_1) = \int_{t_1}^\infty \odif{t} e^{i(E_k - \epsilon E_{p'}) t} \int_{-\infty}^t \odif{t'} e^{i\epsilon E_{p'} t'} (e^{i\omega t'} - e^{-i\omega t'}) e^{-iE_p t'}, 
\end{split}\end{equation}
and 
\begin{equation}\begin{split}
    &f_{\phi, \epsilon} (k, q', p; \omega; t_1) = \int_{t_1}^\infty \odif{t} e^{i(E_k - V_0 - \epsilon E_{q'}) t} \int_{-\infty}^t \odif{t'} e^{i(V_0 + \epsilon E_{q'})t'} (e^{i\omega t'} - e^{-i\omega t'}) e^{-iE_p t'}. 
\end{split}\end{equation} 
These time integrals comprise the exponentially oscillating factors in the scattering wave functions and the vector potential, producing delta functions and Cauchy's principal values. For example, $f_{\psi,+}$ is calculated to be 
\begin{equation}\begin{split}
    &\frac{i}{E_k + \omega - E_p + i0_+} \frac{-i}{E_{p'} + \omega - E_p - i0_+} e^{i(E_k + \omega - E_p) t_1} \\ 
    &- \frac{i}{E_k - \omega - E_p + i0_+} \frac{-i}{E_{p'} - \omega - E_p - i0_+} e^{i(E_k - \omega - E_p) t_1}. 
\end{split}\end{equation}
Taking into account the limit formulae \eqref{eq:(sec3)limit_formula_1}, \eqref{eq:(sec3)limit_formula_2}, we can expect energy conservation relations 
\begin{align}
    E_k = E_{p'} = E_p \pm \omega \label{eq:(sec4)energy_conserv_relation}
\end{align}
at the left infinity $z_L \to -\infty$, which indicates that the transition from the left-incident state $\psi_s^{(E_p)}$ in the initial wave packet into other eigenstates belonging to the energy eigenvalues $E_p \pm \omega$ can occur. Let us compute the case $E_k = E_{p'} = E_p - \omega$ as an example. Here $E_p - \omega > m$ is needed for the on-shell condition. $z_L$-dependent factor $\bar{\psi}_s^{(E_k)} (z_L) \gamma^3 \psi_{s'}^{(E_{p'})} (z_L)$ includes oscillating exponential terms that can survive in the left infinity limit, i.e., the product of the initial waves and that of the reflected waves $\propto \exp [\mp i(k - p') z_L]$. For $p \, (> 0)$ satisfying $E_p + \epsilon \omega > m$, it is convenient to define an assisted momentum 
\begin{align}
    p_\epsilon = \sqrt{(E_p + \epsilon \omega)^2 - m^2} \label{eq:(sec4)p_epsilon}
\end{align}
with its on-shell energy $E_{p_\epsilon} \equiv \sqrt{p_\epsilon^2 + m^2} = E_p + \epsilon \omega$. Then, the following limit evaluation is possible: 
\begin{equation}\begin{split} 
    &\lim_{z_L \to -\infty} \frac{i}{E_k + \omega - E_p + i0_+} \frac{-i}{E_{p'} + \omega - E_p - i0_+} e^{i(k - p') z_L} \\ 
    &= \theta (E_p - \omega - m) (2\pi)^2 \delta (E_k - E_{p_-}) \delta (E_{p'} - E_{p_-}), \label{eq:(sec4)limit_eval}
\end{split}\end{equation} 
which implies that the reflected wave of the left-incident mode with the energy $E_{p_-}$ can generate the reflected charge. Computation of all other terms in the same way results in the reflection probability of the wave packet $P_\Refl$ up to the first order 
\begin{gather} 
    P_\Refl = \bigl| P_\Refl^{(0)} + P_\Refl^{(1)} + \mathcal{O} (\lambda^2) \bigr|, \\ 
    P_\Refl^{(1)} = \Re \biggl\{ 2\lambda \sum_{\epsilon = \pm} (-\epsilon) \int_{E_p + \epsilon \omega > m} \odif{p} g^* (p_\epsilon) g(p) \mathcal{A}_\Refl (p, p_\epsilon) \biggr\}, \label{eq:(sec4)P_refl(1)} 
\end{gather} 
where 
\begin{equation}\begin{split}
    \mathcal{A}_\Refl (p, p_\epsilon) &= |R_\psi (p_\epsilon)|^2 2\pi \frac{E_{p_\epsilon}}{p_\epsilon} \braket{\psi_s^{(E_{p_\epsilon})}, \psi_s^{(E_p)}} \\ 
    &\quad + \theta (E_{p_\epsilon} - V_0 - m) \sqrt{\frac{E_{p_\epsilon}}{E_{q_\epsilon}}} R_\psi^* (p_\epsilon) T_\phi (q_\epsilon) 2\pi \frac{E_{q_\epsilon}}{q_\epsilon} \braket{\phi_s^{(V_0 + E_{q_\epsilon})}, \psi_s^{(E_p)}} \\ 
    &\quad + \theta (V_0 - m - E_{p_\epsilon}) \sqrt{\frac{E_{p_\epsilon}}{E_{q_\epsilon}}} R_\psi^* (p_\epsilon) T_\phi (q_\epsilon) 2\pi \frac{E_{q_\epsilon}}{q_\epsilon} \braket{\phi_s^{(V_0 - E_{q_\epsilon})}, \psi_s^{(E_p)}}. \label{eq:(sec4)A_refl} 
\end{split}\end{equation} 
The angle brackets in eq.~\eqref{eq:(sec4)A_refl} show transitions of the left-incident mode $\psi_s^{(E_p)}$ to other scattering modes due to the energy supply or absorption $\pm \omega$ by the oscillating electric field. The first term on the right-hand side represents the transition to the left-incident modes belonging to the energy eigenvalues $E_{p_\epsilon}$ via the energy conservation relation \eqref{eq:(sec4)energy_conserv_relation}. The second term on the right-hand side represents the transition to the right-incident modes with positive frequencies $V_0 + E_{q_\epsilon} = E_{p_\epsilon}$, and the coefficients $R_\psi^* (p_\epsilon)$ and $T_\phi (q_\epsilon)$ are brought about from the $z_L$-dependent factor $\bar{\psi}_s^{(E_k)} (z_L) \gamma^3 \phi_{s'}^{(V_0 + E_{q'})} (z_L)$ in eq.~\eqref{eq:(sec4)t-integral_Psi(0)bar_gamma_Psi(1)}. The third term on the right-hand side corresponds to the transition to the right-incident modes with negative frequencies $V_0 - E_{q_\epsilon} = E_{p_\epsilon}$, and the reflection and transmission coefficients originate from $\bar{\psi}_s^{(E_k)} (z_L) \gamma^3 \phi_{s'}^{(V_0 - E_{q'})} (z_L)$. The last process can take place according to the energy conservation relation $E_k = V_0 - E_{q'} = E_p + \epsilon \omega$, which leads to an inequality $m < E_p + \epsilon \omega < V_0 - m$, i.e., the assisted energy $E_{p_\epsilon}$ must belong to the Klein region. Thus, the transition into the negative-frequency modes does not occur under the subcritical step. We also comment that the transition into $\psi_{s'}^{(-E_{p'})}$ is not realizable since it requires the condition $E_k = -E_{p'} = E_p \pm \omega$, that never holds.

By a similar calculation, we obtain the transmission probability of the wave packet up to the first order as follows: 
\begin{gather} 
    P_\Trans = \bigl| P_\Trans^{(0)} + P_\Trans^{(1)} + \mathcal{O} (\lambda^2) \bigr|, \\ 
    P_\Trans^{(1)} = \Re \biggl\{ 2\lambda \sum_{\epsilon = \pm} (-\epsilon) \int_{|E_p + \epsilon \omega - V_0| > m} \odif{p} g^* (p_\epsilon) g(p) \mathcal{A}_\Trans (p, p_\epsilon) \biggr\}, \label{eq:(sec4)P_trans(1)}
\end{gather} 
where 
\begin{equation}\begin{split}
    \mathcal{A}_\Trans (p, p_\epsilon) &= \frac{q_\epsilon}{p_\epsilon} |T_\psi (p_\epsilon)|^2 2\pi \frac{E_{p_\epsilon}}{p_\epsilon} \braket{\psi_s^{(E_{p_\epsilon})}, \psi_s^{(E_p)}} \\ 
    &\quad + \theta (E_{p_\epsilon} - V_0 - m) \frac{q_\epsilon}{p_\epsilon} \sqrt{\frac{E_{p_\epsilon}}{E_{q_\epsilon}}} T_\psi^* (p_\epsilon) R_\phi (q_\epsilon) 2\pi \frac{E_{q_\epsilon}}{q_\epsilon} \braket{\phi_s^{(V_0 + E_{q_\epsilon})}, \psi_s^{(E_p)}} \\ 
    &\quad + \theta (V_0 - m - E_{p_\epsilon}) \frac{q_\epsilon}{p_\epsilon} \sqrt{\frac{E_{p_\epsilon}}{E_{q_\epsilon}}} T_\psi^* (p_\epsilon) R_\phi (q_\epsilon) 2\pi \frac{E_{q_\epsilon}}{q_\epsilon} \braket{\phi_s^{(V_0 - E_{q_\epsilon})}, \psi_s^{(E_p)}} \label{eq:(sec4)A_trans}
\end{split}\end{equation} 
Notice that the the weight functions $g^* (p_\epsilon)$ and $g(p)$ in eqs.~\eqref{eq:(sec4)P_refl(1)}, \eqref{eq:(sec4)P_trans(1)} hardly overlap with each other in the momentum space when the Gaussian width $\sigma$ is much smaller than the assistance energy $\omega$. It means that the first-order corrections are attributed to the condition that the initial state is given as a wave packet. On the other hand, when we compute the second-order correction in the next section, it is shown that the oscillating electric field does affect the reflection and transmission probabilities even in the plane-wave limit $\sigma \to 0$, and even under the subcritical step.

Now, let us confirm that the sum of the probabilities is conserved up to the first-order perturbation. To do this, it has to be shown that the first-order correction of the reflection and transmission probabilities have the same absolute value with opposite signs. We first consider the case $|E_{p_\epsilon} - V_0| < m$, where $P_\Refl^{(1)}$ and $P_\Trans^{(1)}$ are given by 
\begin{gather} 
    P_\Refl^{(1)} = \Re \biggl\{ 2\lambda \sum_\epsilon (-\epsilon) \int_{E_p + \epsilon \omega > m} \odif{p} g^* (p_\epsilon) g(p) 2\pi \frac{E_{p_\epsilon}}{p_\epsilon} \braket{\psi_s^{(E_{p_\epsilon})}, \psi_s^{(E_p)}} \biggr\}, \label{eq:(sec4)P_refl(1)=0} \\ 
    P_\Trans^{(1)} = 0. 
\end{gather} 
The transmission probability is not affected by the first-order perturbation because the assisted energy $E_{p_\epsilon}$ belongs to the total-reflection region (ii). Accordingly, the right-hand side of eq.~\eqref{eq:(sec4)P_refl(1)=0} must vanish. It can be shown by making use of a property in terms of the angle brackets 
\begin{align}
    \braket{\psi_1, \psi_2}^* = \braket{\psi_2, \psi_1}. 
\end{align}
In fact, by using the transformation of the integration variable $\odif{p} = \odif{p_+} \frac{E_p}{p} \frac{p_+}{E_{p_+}}$ and by the replacement of the variables $p_+ \to p$ ($\Leftrightarrow p \to p_-$), we can derive 
\begin{align} 
    &\int_{E_{p_+} > m} \odif{p} g^* (p_+) g(p) 2\pi \frac{E_{p_+}}{p_+} \braket{\psi_s^{(E_{p_+})}, \psi_s^{(E_p)}} \\ 
    &= \biggl( \int_{E_{p_-} > m} \odif{p} g^* (p_-) g(p) 2\pi \frac{E_{p_-}}{p_-} \braket{\psi_s^{(E_{p_-})}, \psi_s^{(E_p)}} \biggr)^*, \notag 
\end{align} 
which results in $P_\Refl^{(1)} = 0$. For the other case $|E_{p_\epsilon} - V_0| > m$, the current conservation relations $|R_\psi (p_\epsilon)|^2 + \frac{q_\epsilon}{p_\epsilon} |T_\psi (p_\epsilon)|^2 = 1$ as well as the so-called reciprocity relations 
\begin{align}
    R_\psi^* (p_\epsilon) T_\phi (q_\epsilon) + \frac{q_\epsilon}{p_\epsilon} T_\psi^* (p_\epsilon) R_\phi (q_\epsilon) = 0 \label{eq:(sec4)reciprocity_relation}
\end{align}
can be used to arrive at the result that the sum $P_\Refl^{(1)} + P_\Trans^{(1)}$ is equal to the right-hand side of eq.~\eqref{eq:(sec4)P_refl(1)=0}, which has already been shown to vanish.

\section{\label{Sec5}Second-order perturbation and dynamically assisted Klein tunneling}

As discussed in the previous section, the left-incident mode $\psi_s^{(E_p)}$ in the wave packet is subject to the oscillating electric field, which supplies or absorbs the energy, giving corrections to the reflection and transmission probabilities of the wave packet. However, it has been shown that the first-order correction does not remain in the plane-wave limit $\sigma \to 0$ because the weight functions $g(p)$ and $g(p_\epsilon)$ in the zeroth- and first-order wave packets $\Psi^{(0)}$ and $\Psi^{(1)}$, which are combined to generate the first-order current $J_z^{(1)}$, do not overlap with each other. On the other hand, the second-order current is given by 
\begin{align}
    J_z^{(2)} (x) = \bar{\Psi}^{(1)} (x) \gamma^3 \Psi^{(1)} (x) + 2\Re \bigl\{ \bar{\Psi}^{(0)} (x) \gamma^3 \Psi^{(2)} (x) \bigr\}, \label{eq:(sec5)Jz(2)} 
\end{align}
where 
\begin{equation}\begin{split} 
    \Psi^{(2)} (x) &= (-e)^2 \int \odif[order=2]{x'} S_\Ret (x, x') \gamma^3 A_3 (x') \int \odif[order=2]{x''} S_\Ret (x', x'') \gamma^3 A_3 (x'') \Psi^{(0)} (x''). \label{eq:(sec5)Psi(2)}
\end{split}\end{equation} 
It is expected that the first term on the right-hand side of eq.~\eqref{eq:(sec5)Jz(2)} will include the product of the weight function $g(p_\epsilon)$ and its complex conjugate, which is nonvanishing in the plane-wave limit. We also expect such nonvanishing contributions in the second term, for eq.~\eqref{eq:(sec5)Psi(2)} implies that $\psi_s^{(E_p)}$ in the initial wave packet experiences transition twice by receiving the energy assistance $\pm 2\omega$ or $\omega - \omega = 0$. For the latter case, the second term of $J_z^{(2)}$ will include an absolute square of $g(p)$, which will remain in the plane-wave limit. In the following, we evaluate the second-order correction of the reflection and transmission probabilities in the plane-wave limit (but formally treat the problem in the wave packet formalism as in the previous section). We also restrict ourselves for simplicity to the regime $E_p < V_0 + m$ with the subcritical potential energy $V_0 < 2m$ and the assistance energy $\omega < 2m$.

\begin{figure}[htbp]
    \centering
    \includegraphics[width=0.6\linewidth]{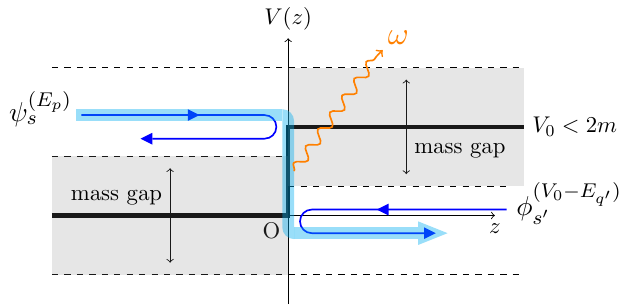}
    \caption{\label{fig:assistedKleinProcess}Schematic picture of the dynamically assisted Klein tunneling under the subcritical potential step and the oscillating field. The thin blue arrow in $z < 0$ represents the positive-frequency left-incident mode $\psi_s^{(E_p)}$ in the initial wave packet, which shows the total reflection. With the aid of the oscillating electric field, the transition to the negative-frequency right-incident mode $\phi_{s'}^{(V_0 - E_{q'})}$ (the thin blue arrow in $z > 0$, which also shows the total reflection) by emitting its energy $\omega$ (the orange wavy line) can occur. Then, the wave packet can reach the right infinity according to the reflected wave of $\phi_{s'}^{(V_0 - E_{q'})}$. Thus, the oscillating electric field enables the tunneling process of the wave packet from the positive-frequency region to the negative-frequency region, as indicated by the sky-blue thick arrow.} 
\end{figure}

Consider the current on the right of the step. If the oscillating electric field is not present, the positive-frequency left-incident mode $\psi_s^{(E_p)}$ belonging to the energy region (ii) totally reflects at $z = 0$. However, now, transitions to other scattering modes can occur by exchanging energy with the oscillating electric field. If the wave packet after the transitions has right-moving oscillating waves on the right side of the step, they will form the transmitted current. Particularly, the first-order wave packet $\Psi^{(1)} (z_R, t)$ has such waves with negative frequencies, which are contained in $\phi_{s'}^{(V_0 - E_{q'})} (z_R, t)$ as their reflected waves, i.e., 
\begin{equation}\begin{split} 
    &\Psi^{(1)} (z_R, t) \\ 
    &\supset \lambda \int_0^\infty \odif{p} g(p) \sum_{s'} \int_0^\infty \odif{q'} \phi_{s'}^{(V_0 - E_{q'})} (z_R, t) \braket{\phi_{s'}^{(V_0 - E_{q'})}, \psi_s^{(E_p)}} \int_{-\infty}^t \odif{t'} e^{i(V_0 - E_{q'} + \omega - E_p) t'}. \label{eq:(sec5)Psi(1)(z_R,t)}
\end{split}\end{equation} 
Thus, the positive-frequency wave incoming from the left can tunnel into the negative-frequency region on the right of the step by emitting its energy to the oscillating electric field and can reach the right infinity. In this paper, we refer to the process as dynamically assisted Klein tunneling; see also its schematic picture in figure~\ref{fig:assistedKleinProcess}.

We confirm it in practice: integrating the current of the right-hand side of eq.~\eqref{eq:(sec5)Psi(1)(z_R,t)} over time reads 
\begin{equation}\begin{split}
    &\int_{t_1}^\infty \odif{t} \bar{\Psi}^{(1)} (z_R, t) \gamma^3 \Psi^{(1)} (z_R, t) \\ 
    &\supset \lambda^2 \int_0^\infty \odif{k} \odif{p} g^* (k) g(p) \sum_{s', s''} \int_0^\infty \odif{q'} \odif{q''} \\ 
    &\quad \times \braket{\phi_{s''}^{(V_0 - E_{q''})}, \psi_s^{(E_k)}}^* \braket{\phi_{s'}^{(V_0 - E_{q'})}, \psi_s^{(E_p)}} \bar{\phi}_{s''}^{(V_0 - E_{q''})} (z_R) \gamma^3 \phi_{s'}^{(V_0 - E_{q'})} (z_R) \\ 
    &\quad \times \frac{-i}{E_k - \omega - V_0 + E_{q''} - i0_+} \frac{-i}{V_0 - E_{q'} + \omega - E_p - i0_+} \frac{i}{E_k - E_p + i0_+} e^{i(E_k - E_p) t_1}. 
\end{split}\end{equation} 
As in eq.~\eqref{eq:(sec4)t-integral_Psi(0)bar_gamma_Psi(1)}, $k$ and $p$ denote the momenta of the zeroth-order wave packet included in $\bar{\Psi}^{(1)} (z_R, t)$ and $\Psi^{(1)} (z_R, t)$, while $q'$, $q''$, $s'$, and $s''$ are the momenta and spin that are brought about from the Green function. The $z_R$-dependent factor $\bar{\phi}_{s''}^{(V_0 - E_{q''})} (z_R) \gamma^3 \phi_{s'}^{(V_0 - E_{q'})} (z_R)$ on the above expression includes an oscillating term $\propto \exp [i(q'' - q') z_R]$, which is a product of the reflected waves. Notice that for $p$ satisfying $V_0 - E_p + \omega > m$, an assisted momentum 
\begin{align}
    q_- = \sqrt{(V_0 - E_p + \omega)^2 - m^2} \label{eq:(sec5)q_minus}
\end{align}
exists, with its on-shell energy $E_{q_-} = \sqrt{q_-^2 + m^2} = V_0 - E_p + \omega$. Then, we can perform a similar calculation as eq.~\eqref{eq:(sec4)limit_eval} by using the limit formulae in the previous section to obtain 
\begin{equation}\begin{split} 
    &\lim_{z_R \to \infty} \frac{-i}{E_k - \omega - V_0 + E_{q''} - i0_+} \frac{-i}{V_0 - E_{q'} + \omega - E_p - i0_+} \frac{i}{E_k - E_p + i0_+} e^{i(q'' - q') z_R} \\ 
    &= \theta (V_0 - E_p + \omega - m) (2\pi)^3 \delta (E_{q'} - E_{q_-}) \delta (E_{q''} - E_{q_-}) \delta (E_k - E_p). 
\end{split}\end{equation} 
It implies that the reflected wave of the negative-frequency mode $\phi_{s'}^{(V_0 - E_{q'})} (z_R)$ with the momentum $q' = q_-$ in the right-hand side of eq.~\eqref{eq:(sec5)Psi(1)(z_R,t)} does in fact contribute to the tunneling current.

As the unperturbed wave packet with the energy $E_p < V_0 + m$ cannot reach the right infinity, the second term of the right-hand side of $J_z^{(2)}$ in eq.~\eqref{eq:(sec5)Jz(2)} does not contribute to the second-order correction of the transmission probability. Thus, we arrive at the following expression: 
\begin{equation}\begin{split}
    P_\Trans^{(2)} &= \lambda^2 \int_{E_p + \omega > V_0 + m} \odif{p} |g(p)|^2 \mathcal{B}_\Trans (p, p_+) + \lambda^2 \int_{E_p - \omega < V_0 - m} \odif{p} |g(p)|^2 \mathcal{C} (p, q_-), \label{eq:(sec5)P_trans(2)}
\end{split}\end{equation}
where $\mathcal{B}_\Trans$ and $\mathcal{C}$ are given by 
\begin{equation}\begin{split}
    &\mathcal{B}_\Trans (p, p_+) \\ 
    &= \frac{q_+}{p_+} |T_\psi (p_+)|^2 (2\pi)^2 \frac{E_p}{p} \frac{E_{p_+}}{p_+} \sum_{s'} \bigl| \braket{\psi_{s'}^{(E_{p_+})}, \psi_s^{(E_p)}} \bigr|^2 \\ 
    &\quad + |R_\phi (q_+)|^2 (2\pi)^2 \frac{E_p}{p} \frac{E_{q_+}}{q_+} \sum_{s'} \bigl| \braket{\phi_{s'}^{(V_0 + E_{q_+})}, \psi_s^{(E_p)}} \bigr|^2 \\ 
    &\quad + 2\Re \biggl\{ \sqrt{\frac{E_{q_+}}{E_{p_+}}} T_\psi^* (p_+) R_\phi (q_+) (2\pi)^2 \frac{E_p}{p} \frac{E_{p_+}}{p_+} \sum_{s'} \braket{\psi_{s'}^{(E_{p_+})}, \psi_s^{(E_p)}}^* \braket{\phi_{s'}^{(V_0 + E_{q_+})}, \psi_s^{(E_p)}} \biggr\} 
\end{split}\end{equation}
and 
\begin{align}
    \mathcal{C} (p, q_-) = (2\pi)^2 \frac{E_p}{p} \frac{E_{q_-}}{q_-} \sum_{s'} \bigl| \braket{\phi_{s'}^{(V_0 - E_{q_-})}, \psi_s^{(E_p)}} \bigr|^2, \label{eq:(sec5)cal_C}
\end{align} 
respectively. $\mathcal{B}_\Trans$ corresponds to the over-the-barrier scattering processes which are made possible by absorbing the energy $+\omega$ from the oscillating electric field. The first and second terms on the right-hand side represent the transitions to the modes $\psi_{s'}^{(E_{p_+})}$ and $\phi_{s'}^{(V_0 + E_{q_+})}$. The third term on the right-hand side is due to the degeneracy of $\psi_{s'}^{(E_{p_+})}$ and $\phi_{s'}^{(V_0 + E_{q_+})}$, where $E_{p_+} = V_0 + E_{q_+}$. On the other hand, $\mathcal{C}$ represents the contribution arising from the dynamically assisted Klein tunneling. We stress that the mode functions $\psi_s^{(E_p)}$ and $\phi_{s'}^{(V_0 - E_{q_-})}$ in the angle-bracket form in eq.~\eqref{eq:(sec5)cal_C} have an overlap around $z \sim 0$ even though they show the total reflection on the left and right of the step, respectively. Thus, it has been shown that the tunneling from the positive-frequency region to the negative-frequency region can take place at the second order of the Furry-picture perturbation even when the potential energy $V_0$ as well as the assistance energy $\omega$ are subcritical. What is significant in the above discussion is the existence of the momentum of the negative-frequency wave; $q_-$ introduced in eq.~\eqref{eq:(sec5)q_minus}. It implies that the following inequality for the potential energy $V_0$ and the assistance energy $\omega$ 
\begin{align}
    V_0 + \omega > 2m \label{eq:(sec5)criticality_condition}
\end{align}
must be satisfied as a necessary condition. It is nothing but the supercritical condition for the dynamically assisted Klein tunneling.

We calculate the second-order correction of the reflection probability $P_\Refl^{(2)}$ for completeness. As discussed at the beginning of section~\ref{Sec3}, the probability conservation relation holds independent of perturbation, which implies $P_\Refl^{(2)} = -P_\Trans^{(2)}$. In particular, a scattering process should exist that decreases the reflection probability in accordance with the increase of the transmission probability due to the dynamically assisted Klein tunneling. Such a process is that the incoming wave emits its energy $\omega$ once to penetrate the negative-frequency band below the step, and then it gains $\omega$ and returns to the positive-frequency band, moving away to the left infinity, i.e.,  
\begin{equation}\begin{split} 
    &\Psi^{(2)} (z_L, t) \\ 
    &\supset -\lambda^2 \int_0^\infty \odif{p} g(p) \sum_{s', s''} \int_0^\infty \odif{q'} \odif{p''} \psi_{s''}^{(E_{p''})} (z_L, t) \braket{\psi_{s''}^{(E_{p''})}, \phi_{s'}^{(V_0 - E_{q'})}} \braket{\phi_{s'}^{(V_0 - E_{q'})}, \psi_s^{(E_p)}} \\ 
    &\quad \times \int_{-\infty}^t \odif{t'} e^{i(E_{p''} - \omega - V_0 + E_{q'}) t'} \int_{-\infty}^{t'} \odif{t''} e^{i(V_0 - E_{q'} + \omega - E_p) t''}. \label{eq:(sec5)Psi(2)(z_L,t)} 
\end{split}\end{equation} 
The first half of the process is characterized by the angle bracket $\braket{\phi_{s'}^{(V_0 - E_{q'})}, \psi_s^{(E_p)}}$, and the latter half is $\braket{\psi_{s''}^{(E_{p''})}, \phi_{s'}^{(V_0 - E_{q'})}}$. Calculating the time integral of the current composed of $\Psi^{(0)}$ and the right-hand side of eq.~\eqref{eq:(sec5)Psi(2)(z_L,t)} leads to 
\begin{equation}\begin{split} 
    &\int_{t_1}^\infty \odif{t} \bar{\Psi}^{(0)} (z_L, t) \gamma^3 \Psi^{(2)} (z_L, t) \\ 
    &\supset -\lambda^2 \int_0^\infty \odif{k} \odif{p} g^* (k) g(p) \sum_{s', s''} \int_0^\infty \odif{q'} \odif{p''} \\ 
    &\quad \times \braket{\psi_{s''}^{(E_{p''})}, \phi_{s'}^{(V_0 - E_{q'})}} \braket{\phi_{s'}^{(V_0 - E_{q'})}, \psi_s^{(E_p)}} \bar{\psi}_s^{(E_k)} (z_L) \gamma^3 \psi_{s''}^{(E_{p''})} (z_L) \\ 
    &\quad \times \frac{-i}{E_{p''} - E_p - i0_+} \frac{-i}{V_0 - E_{q'} + \omega - E_p - i0_+} \frac{i}{E_k - E_p + i0_+} e^{i(E_k - E_p) t_1}. \label{eq:(sec5)integral_current(0)(2)}
\end{split}\end{equation} 
The $z_L$-dependent factor $\bar{\psi}_s^{(E_k)} (z_L) \gamma^3 \psi_{s''}^{(E_{p''})} (z_L)$ includes an oscillating exponential $\exp [i(k - p'') z_L]$ brought about from the reflected waves, with which we evaluate the limit $z_L \to -\infty$ to obtain 
\begin{equation}\begin{split} 
    &\lim_{z_L \to -\infty} \frac{-i}{E_{p''} - E_p - i0_+} \frac{-i}{V_0 - E_{q'} + \omega - E_p - i0_+} \frac{i}{E_k - E_p + i0_+} e^{i(k - p'') z_L} \\ 
    &= \frac{-i}{V_0 - E_{q'} + \omega - E_p - i0_+} (2\pi)^3 \delta (E_{p''} - E_p) \delta (E_k - E_p). \label{eq:(sec5)q'_unconstrained}
\end{split}\end{equation} 
Note that the limit does not give any constraints for the intermediate momentum $q'$, implying that when substituted to the right-hand side of eq.~\eqref{eq:(sec5)integral_current(0)(2)}, not only the $p$-integration but also the $q'$-integration remains. However, we must remind here that only the real part of eq.~\eqref{eq:(sec5)integral_current(0)(2)} contributes to the second-order current. The distribution $-i/(V_0 - E_{q'} + \omega - E_p - i0_+)$ on the right-hand side of eq.~\eqref{eq:(sec5)q'_unconstrained} consists of the delta function and the Cauchy principal value. Observe that the principal value produces just an imaginary part of eq.~\eqref{eq:(sec5)integral_current(0)(2)}, which is irrelevant to the current under discussion. It means that only the delta-function part, which contracts the $q'$-integration, contributes to the current. Therefore, we arrive at the result of the second-order correction of the reflection probability as follows: 
\begin{equation}\begin{split}
    P_\Refl^{(2)} &= - \lambda^2 \int_{E_p + \omega > V_0 + m} \odif{p} |g(p)|^2 \mathcal{B}_\Refl (p, p_+) - \lambda^2 \int_{E_p - \omega < V_0 - m} \odif{p} |g(p)|^2 \mathcal{C} (p, q_-), \label{eq:(sec5)P_refl(2)}
\end{split}\end{equation}
where $\mathcal{B}_\Refl$ is a contribution of the over-the-barrier scattering with the energy assistance $+\omega$: 
\begin{equation}\begin{split}
    &\mathcal{B}_\Refl (p, p_+) \\ 
    &= (1 - |R_\psi (p_+)|^2) (2\pi)^2 \frac{E_p}{p} \frac{E_{p_+}}{p_+} \sum_{s'} \bigl| \braket{\psi_{s'}^{(E_{p_+})}, \psi_s^{(E_p)}} \bigr|^2 \\ 
    &\quad + |R_\phi (q_+)|^2 (2\pi)^2 \frac{E_p}{p} \frac{E_{q_+}}{q_+} \sum_{s'} \bigl| \braket{\phi_{s'}^{(V_0 + E_{q_+})}, \psi_s^{(E_p)}} \bigr|^2 \\ 
    &\quad - 2\Re \biggl\{ \frac{p_+}{q_+} \sqrt{\frac{E_{q_+}}{E_{p_+}}} R_\psi^* (p_+) T_\phi (q_+) (2\pi)^2 \frac{E_p}{p} \frac{E_{p_+}}{p_+} \sum_{s'} \braket{\psi_{s'}^{(E_{p_+})}, \psi_s^{(E_p)}}^* \braket{\phi_{s'}^{(V_0 + E_{q_+})}, \psi_s^{(E_p)}} \biggr\}, 
\end{split}\end{equation} 
while $\mathcal{C}$ has been given by eq.~\eqref{eq:(sec5)cal_C}. By utilizing the reciprocity relation \eqref{eq:(sec4)reciprocity_relation}, we can easily check that 
\begin{align}
    P_\Refl^{(2)} = -P_\Trans^{(2)}. 
\end{align}

In the remainder of this section, we show the results of the above analytical calculations and compare them with numerical simulations (the settings of the numerical simulations are given in the appendix). Figure~\ref{fig:Ptrans(V0=1.5)} shows $\omega$-dependence of the transmission probability $P_\Trans$ for the central momentum $\bar{p}/m = 1.0$ and the potential height $V_0/m = 1.5$. The intensity and spatial width of the oscillating electric field are $e\mathcal{E}_z/m^2 = 0.04$, $l/\lambdabar = 1.0$, where $\lambdabar = \hbar/(mc)$ is the Compton wavelength. Here, the perturbation parameter $\lambda = e\mathcal{E}_z/(2 m \omega)$ varies from $2/3 \times 10^{-2}$ to $4 \times 10^{-2}$. The momentum deviation of the wave packet is set to $\sigma/m = 0.1$. If $\omega$ is sufficiently larger than the momentum deviation, the second-order perturbation computed in the plane-wave limit $\sigma \to 0$, discussed above, is expected to be a good approximation. 
\begin{figure}[htbp]
    \centering
    \includegraphics[width=0.8\linewidth]{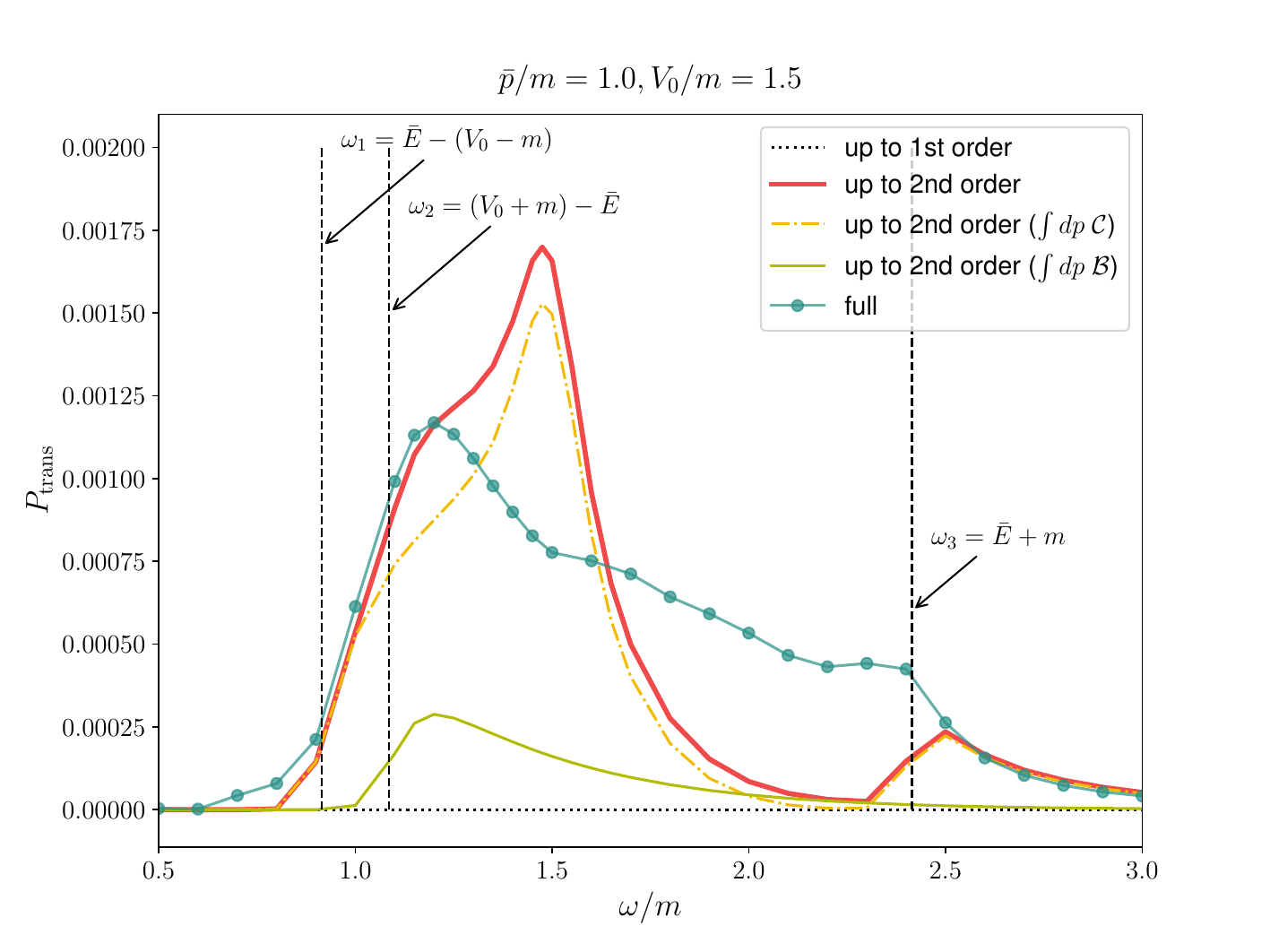}
    \caption{\label{fig:Ptrans(V0=1.5)}Transmission probability of the wave packet $P_\Trans$ for the central momentum $\bar{p}/m = 1.0$ and the potential height $V_0/m = 1.5$. Other parameters are $\sigma/m = 0.1$, $e\mathcal{E}_z/m^2 = 0.04$, and $l/\lambdabar = 1.0$, where $\lambdabar$ is the Compton wavelength. The analytical results up to the first and second orders are shown in the black dotted line and the thick red line, respectively. The yellow dash-dotted line (the light green solid line) represents the analytical result up to the second order except for the first (second) term on the right-hand side of $P_\Trans^{(2)}$ in eq.~\eqref{eq:(sec5)P_refl(2)}. The numerical result is plotted in blue-green.} 
\end{figure}
Since the on-shell energy corresponding to the central momentum is $\bar{E} \equiv E_{\bar{p}} = \sqrt{2}m < V_0 + m$, most of the wave packet should be reflected without energy assistance. As $\omega$ increases from zero, the transmission probability gradually increases, and a significant increase is seen around $\omega_1 = \bar{E} - (V_0 - m)$, beyond which the central momentum mode can cause dynamically assisted Klein tunneling. More precisely, the transmission probability begins to increase before $\omega$ exceeds $\omega_1$ because we consider the scattering of the wave packet. We comment that for small $\omega/m \ll 1$, the incident wave packet hardly transmits the potential barrier if the assistance energy is too small, which causes the suppression of the transmission probability. Furthermore, above $\omega_2 = (V_0 + m) - \bar{E}$, the assisted over-the-barrier scattering of the central momentum mode also contributes to the transmission probability; in fact, the numerical result (blue-green plots) representing its contribution begins to rise near the value $\omega_2$. It is observed that below $\omega/m \lesssim 1.2$, the analytical result up to the second-order perturbation (thick red line) is in good agreement with the numerical result, where the contribution of the assisted Klein tunneling (yellow dash-dotted line) is dominant. Above $\omega/m = 1.2$, we see the breakdown of the perturbation approximation. However, when $\omega$ exceeds $\omega_3 = \bar{E} + m$, where the oscillating electric field can affect the reflection and transmission probabilities in the absence of the potential step, the analytical result reproduces the numerical result well, although we have neglected the contribution of $\omega > 2m$ in the analytical discussion. Here, the perturbation parameter $\lambda$ is sufficiently small that the perturbation approximation is considered to work well.

\begin{figure}[htbp]
    \centering
    \includegraphics[width=0.7\linewidth]{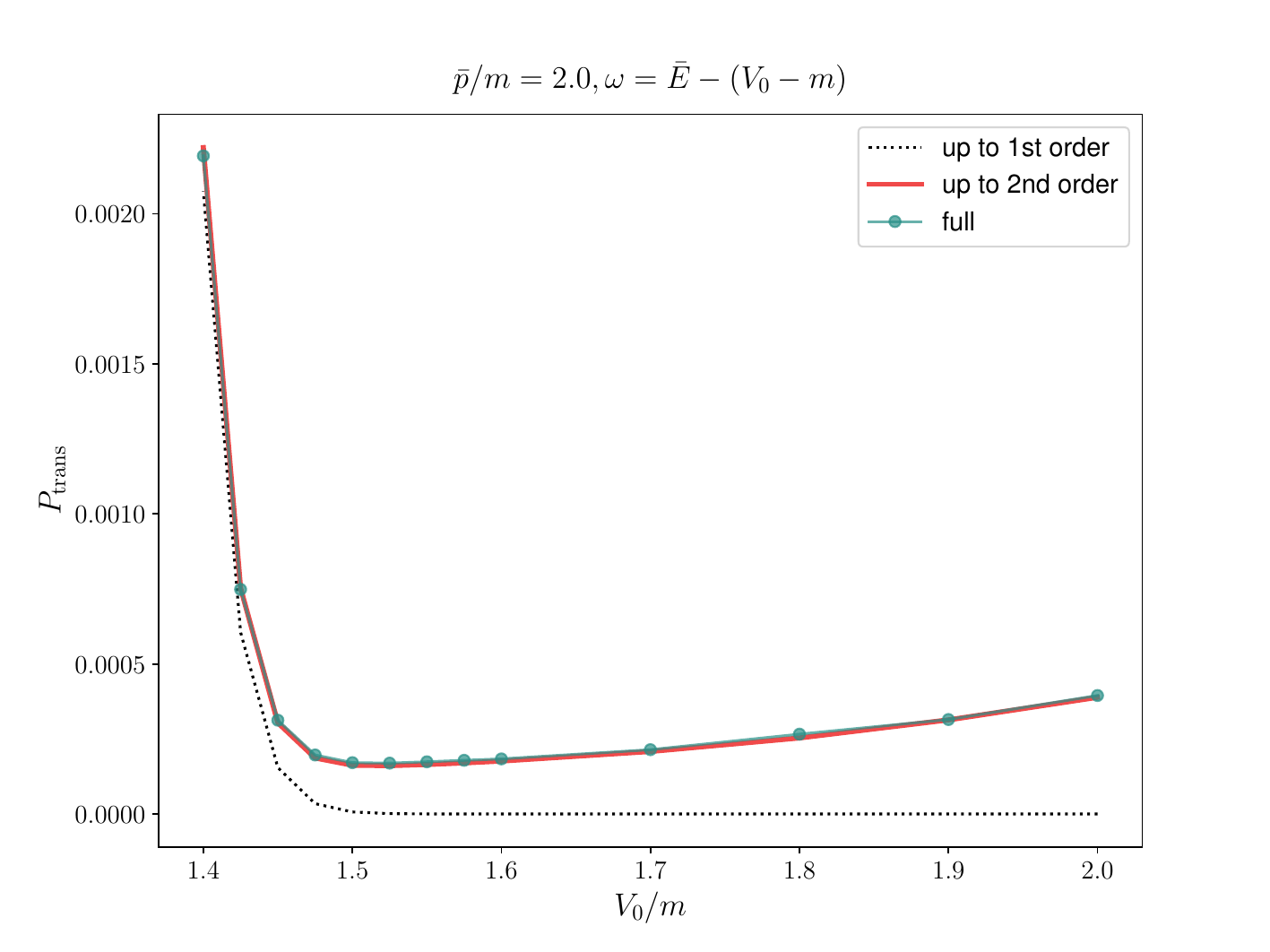}
    \caption{\label{fig:Ptrans(omega=Ebar-V0+m)}Transmission probability of the wave packet $P_\Trans$ for the central momentum $\bar{p}/m = 2.0$ and the assistance energy $\omega = \bar{E}-(V_0 - m)$. Other parameters are $\sigma/m = 0.1$, $e\mathcal{E}_z/m^2 = 0.04$, and $l/\lambdabar = 1.0$. The numerical result is shown by the blue-green plots, while the analytical results up to the first and second orders are represented by the black dotted line and the thick red line, respectively.} 
\end{figure}

We also show $V_0$-dependence of the transmission probability when the assistance energy is $\omega = \omega_1$, where the dynamically assisted Klein tunneling begins to occur (see figure~\ref{fig:Ptrans(omega=Ebar-V0+m)}). The central momentum is $\bar{p}/m = 2.0$, and the other parameters are the same as those in the previous figure. 
It is seen that the transmission probability up to the second order (thick red line) agrees very well with the full result (blue-green plots) for $1.4 \leq V_0/m \leq 2.0$. This result supports the validity of the mechanism of dynamically assisted Klein tunneling based on the Furry picture, which is described at the beginning of the section and figure~\ref{fig:assistedKleinProcess}.

\section{\label{Sec6}Conclusion and Future work}

In this study, we have discussed the one-dimensional scattering of a relativistic fermion in the presence of both the potential step and the oscillating electric field within relativistic quantum mechanics. Since the framework of stationary scattering cannot be applied to the case of time-dependent backgrounds, the wave packet formalism is developed. We calculate in section~\ref{Sec3} the reflection and transmission probabilities of the wave packet \eqref{eq:(sec3)P_refl_def} and \eqref{eq:(sec3)P_refl_def} under the potential step alone to derive consistent results with those in the conventional stationary scattering. The definition of these probabilities is based on the continuity relation for the probability density and current density of the wave packet, which holds independently of the configuration of external fields. The key ingredients in the calculation are the Riemann--Lebesgue lemma as well as the limit formulae in eqs.~\eqref{eq:(sec3)limit_formula_1}, \eqref{eq:(sec3)limit_formula_2}. As a result, the reflection and transmission probabilities, which in the plane-wave limit of the wave packet coincide with those in the stationary scattering, have been obtained. In section~\ref{Sec4} and section~\ref{Sec5}, the reflection and transmission probabilities of the wave packet in the presence of the oscillating electric field superimposed on the step potential are calculated up to the first- and second-order perturbation, respectively. We utilize the Furry-picture perturbation theory, where the oscillating electric field is treated perturbatively while the step potential is non-perturbatively. As the retarded Green function \eqref{eq:(sec4)S_ret} carries not only the perturbative contribution from the oscillating electric field but also the non-perturbative contribution from the potential step, the perturbation theory produces nontrivial results, including the coorperative effects of these contributions. In particular, we demonstrate that the tunneling process between the positive- and negative-frequency regions can occur at the second-order perturbation, even in the absence of the Klein region, which we refer to as dynamically assisted Klein tunneling. The most significant point in the process is the transition between the positive-frequency left-incident mode $\psi_s^{(E_p)}$ and the negative-frequency right-incident mode $\phi_{s'}^{(V_0 - E_{q'})}$ by exchanging energy $\omega$ to the oscillating electric field, depicted in figure~\ref{fig:assistedKleinProcess}. Here, the supercritical condition for the tunneling is altered from that for conventional Klein tunneling, i.e., $V_0 + \omega > 2m$. This means that dynamically assisted Klein tunneling can occur if the stationary potential and the oscillating electric field are imposed simultaneously.

We have only treated the simplest case of the relativistic scattering problem in which temporal potentials are imposed along with a stationary potential. Furry-picture perturbation theory, adopted in this paper, is also applicable for other potential configurations, such as an electric field that is not periodic in time. It is also interesting to consider a stationary potential with a more complicated profile, such as the tangent-hyperbolic potential (also known as the Sauter potential~\cite{sauter_1932_Z.Phys._KleinschenParadoxon, chervyakov.kleinert_2009_Phys.Rev.D_ExactPairProduction, chervyakov.kleinert_2018_Phys.Part.Nuclei_ElectronPositronPair}) or the square-barrier potential~\cite{dombey.calogeracos_1999_Phys.Rep._SeventyYearsKlein, calogeracos.dombey_1999_Contemp.Phys._HistoryPhysicsKlein}. For the latter potential, it may be an interesting problem to investigate whether bound states, which do not exist under the step case, can affect dynamically assisted Klein tunneling.

Klein tunneling has also been discussed in the context of condensed matter physics, such as graphene, where the carriers show a relativistic energy-momentum dispersion relation due to symmetries of the crystal structure. Graphene is highly expected as one of the future semiconductor materials, and its various electric transport properties, including Klein tunneling, have been actively investigated~\cite{katsnelson.novoselov.ea_2006_Nat.Phys._ChiralTunnellingKlein, bai.zhang_2007_Phys.Rev.B_KleinParadoxResonant, pereira.peeters.ea_2010_Semicond.Sci.Technol._KleinTunnelingSingle, allain.fuchs_2011_Eur.Phys.J.B_KleinTunnelingGraphene, navarro-giraldo.quimbay_2020_Ann.Phys._TwoDimensionalKleinTunneling, schmitt.vallet.ea_2023_NaturePhysics_MesoscopicKleinSchwingerEffect}. We have discussed dynamically assisted Klein tunneling only in the one-dimensional case; however, by generalizing the framework of the wave packet scattering, this phenomenon may also be discussed for two-dimensional graphene systems.

\appendix 
\section{\label{AppendA}Klein tunneling, Klein paradox, and numerical simulations}

\begin{figure}[htbp]\centering
  \begin{minipage}{0.5\columnwidth}\centering 
      \includegraphics[width=0.8\columnwidth]{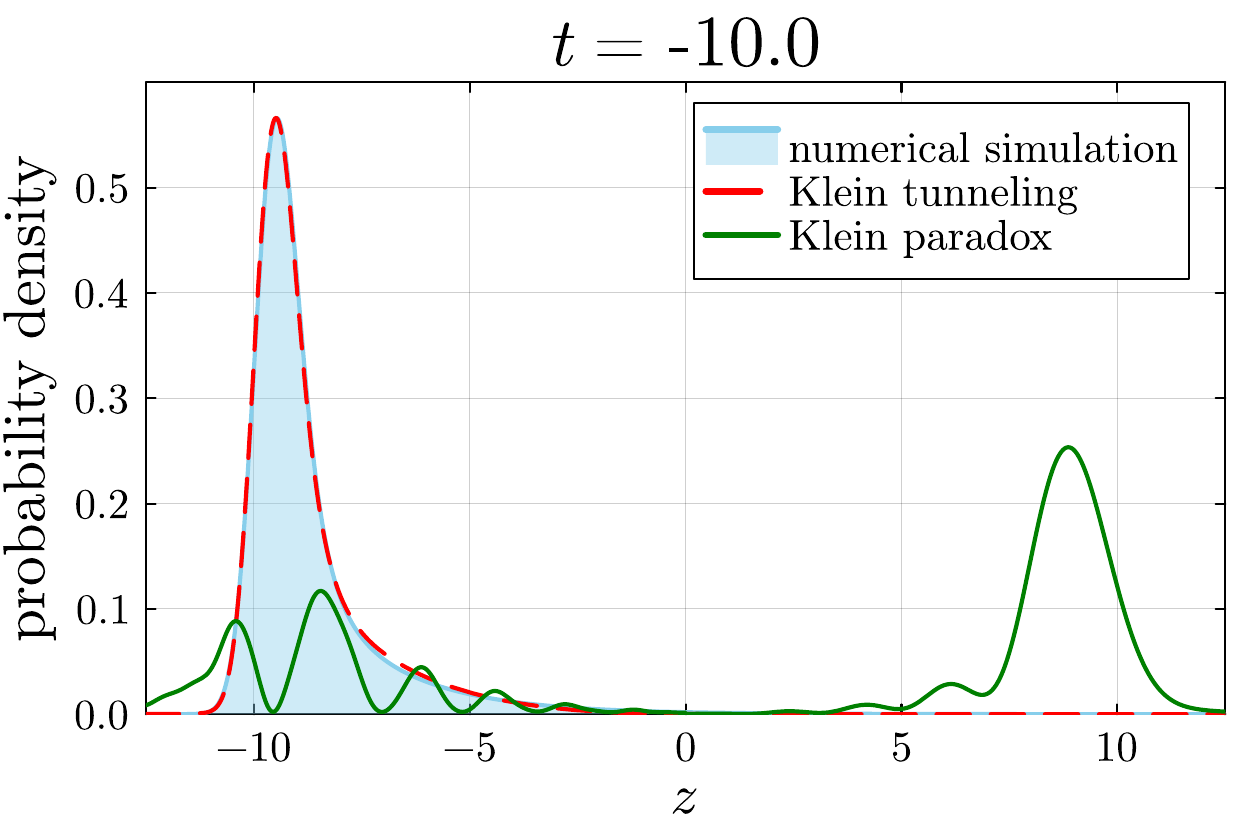}
  \end{minipage}
  \begin{minipage}{0.5\columnwidth}\centering 
      \includegraphics[width=0.8\columnwidth]{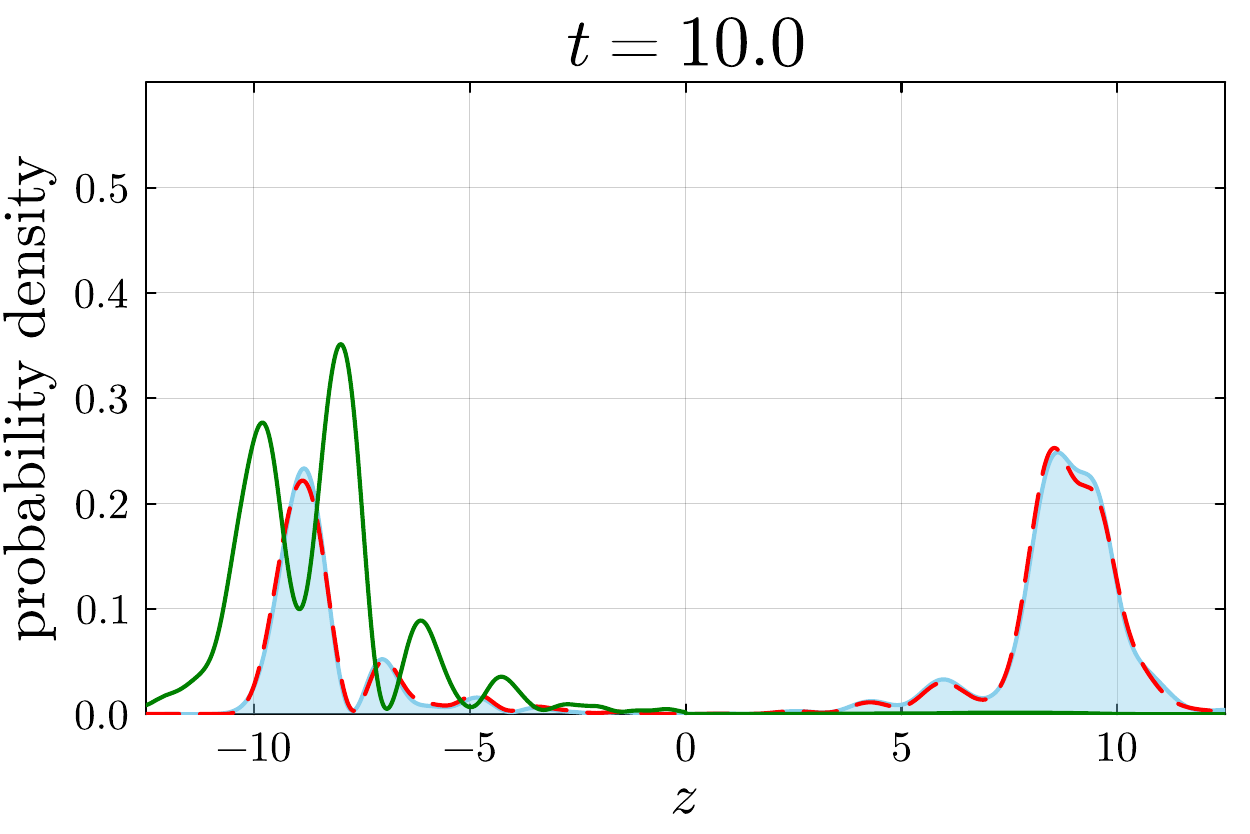}
  \end{minipage}
  \caption{\label{fig:KleinTunnelingParadox}The probability densities of the wave packets under the potential step with height $V_0 = 5.0$ at times $t = -10.0$ (left panel) and $t = 10.0$ (right panel), in units of $m = 1$. The red dashed line and the green solid line correspond to Klein tunneling and Klein paradox, respectively. The sky-blue region is the numerical simulation.} 
\end{figure}

As seen in eq.~\eqref{eq:(sec3)psi_KleinTunneling}, Klein tunneling indicates the solution in which the transmitted wave has a positive group velocity along the $z$-direction. Thus, if we construct the wave packet using the solution such as in eq.~\eqref{eq:(sec3)Psi0}, it is initially incident from the left infinity and hits the step potential at $z = 0$. The reflected and transmitted wave packets then move to the left and right infinity, respectively. Klein paradox, on the other hand, refers to a solution in which the transmitted momentum flips its sign, i.e., 
\begin{align}
    \tilde{\psi}_s^{(E_p)} (z) &\propto \theta (-z) \bigl[ u(p, s) e^{ipz} + \tilde{R} (p) u(-p, s) e^{-ipz} \bigr] + \theta (z) \tilde{T} (p) v(-q, s) e^{-iqz}. \label{eq:(appendA)psi_KleinParadox}
\end{align}
Here, $q \, (> 0)$ is determined by the same relation $E_p = V_0 - E_q$ as that in the Klein tunneling solution \eqref{eq:(sec3)psi_KleinTunneling}. Two coefficients $\tilde{R} (p)$ and $\tilde{T} (p)$ satisfy 
\begin{align}
    |\tilde{R} (p)|^2 - \frac{q}{p} |\tilde{T} (p)|^2 = 1, 
\end{align}
which gives the paradoxical implication that the reflection probability is greater than one. However, since the group velocity of the transmitted wave is negative, the wave packet constructed by using the Klein paradox solution does not describe a physically adequate scattering process. Figure~\ref{fig:KleinTunnelingParadox} shows the probability densities of these two types of wave packets under the overcritical potential step, which are compared to the numerical simulation. The initial state of the numerical simulation is given by 
\begin{align} 
  \Psi_\mathrm{num}^{(0)} (z, t_\mathrm{i}) = \int_0^\infty \odif{p} g(p) u_{p, s} (z, t_\mathrm{i}), 
\end{align} 
where $t_\mathrm{i} = -10.0\lambdabar/c$ is the initial time and $u_{p, s} (z, t)$ is a positive-frequency plane wave 
\begin{align} 
  u_{p, s} (z, t) = \frac{1}{\sqrt{2\pi}} \sqrt{\frac{m}{E_p}} u(p, s) e^{-iE_p t + ipz}. 
\end{align} 
We have applied the FFT-split operator method up to the second order~\cite{braun.su.ea_1999_Phys.Rev.A_NumericalApproachSolve, mocken.keitel_2004_J.Comput.Phys._QuantumDynamicsRelativistic, mocken.keitel_2008_Comput.Phys.Commun._FFTsplitoperatorCodeSolving}. Parameters in units of $m = 1$ are as follows: spatial length $L = 100.0$, total scattering time $T = 20.0$, position $z \in [-L/2, L/2]$, time $t \in [-T/2, T/2]$, numbers of divisions of space and time $N_z = 4000$, $N_t = 2000$ (We have adopted the same parameter settings to produce the numerical results in figure~\ref{fig:Ptrans(V0=1.5)} and~\ref{fig:Ptrans(omega=Ebar-V0+m)}). 
All the probability densities are normalized to one by the normalization condition \eqref{eq:(sec3)normalization_rho(0)}. We can see that in the case of Klein tunneling, the probability density of the wave packet is initially localized on the left of the step, and it splits into two parts after the collision to $z = 0$. This behavior is in good agreement with the numerical simulation. In the case of Klein paradox, on the other hand, the probability density of the wave packet already distributes on each side of the step from the beginning. They are then merged into one at $z = 0$ and localized on the left of the step, as if the time order were reversed. 

\section{ \label{AppendB} Limit formulae for distributions }

Here, we show the derivation of eqs.~\eqref{eq:(sec3)limit_formula_1} and \eqref{eq:(sec3)limit_formula_2}, which play an essential role in evaluating the reflection and transmission probabilities of the wave packet. First, we have to keep in mind the Sokhotski--Plemelj formula 
\begin{align}
\frac{\pm i}{k \pm i0_+} = \pi\delta(k) \pm i PV \frac{1}{k}, 
\end{align}
where $PV$ denotes the principal part. Given a regular function $f(k)$ on the Fourier space, let us consider the limit $z \to -\infty$ of the following principal-value integral 
\begin{align}
    \lim_{z \to -\infty} PV \int_{-\infty}^{\infty} \odif{k} \frac{f(k)}{k} e^{ik z}. 
\end{align}
If the integrand were a well-behaved function, the Fourier integral would vanish in the limit due to the Riemann--Lebesgue lemma. Now, the integrand has a singularity at $k = 0$, which produces a nonvanishing contribution to the above limit. In fact, we can evaluate it by changing the integration variable $k' = kz$, as 
\begin{equation*}\begin{split}
PV \int_{-\infty}^{\infty} \odif{k} \frac{f(k)}{k} e^{ik z}
&= -PV \int_{-\infty}^{\infty} dk' \frac{f(k'/z)}{k'} e^{ik'} \\
&\xrightarrow{z \to -\infty} - f(0) PV \int_{-\infty}^{\infty} dk' \frac{1}{k'} e^{ik'} = - i\pi f(0).
\end{split}\end{equation*}
Therefore, we obtain a formula 
\begin{align}
\lim_{z\to-\infty} PV \frac{1}{k} e^{ik z} = -i\pi \delta(k). 
\end{align}
By using the relations 
\begin{align}
\frac{1}{E_p - E_{p'}} = \frac{1}{p-p'} \frac{E_p + E_{p'}}{p + p'}, \quad 
\delta(E_p - E_{p'}) = \frac{E_p}{p} \delta(p-p'), 
\end{align}
we arrive at the limit formulae \eqref{eq:(sec3)limit_formula_1} and \eqref{eq:(sec3)limit_formula_2}. 

\begin{acknowledgments}
The authors would like to thank H. Nakazato for reading the manuscript and providing numerous critical and helpful comments. 
\end{acknowledgments}

\bibliographystyle{JHEP}
\bibliography{refs}

\providecommand{\href}[2]{#2}\begingroup\raggedright\begin{thebibliography}{10}

\bibitem{klein_1929_Z.Phys._ReflexionElektronenPotentialsprung}
O.~Klein, \emph{Die {Reflexion} von {Elektronen} an einem {Potentialsprung} nach der {Relativistischen} {Dynamik} von {Dirac}}, \href{https://doi.org/10.1007/BF01339716}{\emph{Z. Phys.} {\bfseries 53} (1929) 157}.

\bibitem{dombey.calogeracos_1999_Phys.Rep._SeventyYearsKlein}
N.~Dombey and A.~Calogeracos, \emph{Seventy {Years} of the {Klein} {Paradox}}, \href{https://doi.org/10.1016/S0370-1573(99)00023-X}{\emph{Phys. Rep.} {\bfseries 315} (1999) 41}.

\bibitem{calogeracos.dombey_1999_Contemp.Phys._HistoryPhysicsKlein}
A.~Calogeracos and N.~Dombey, \emph{History and {Physics} of the {Klein} {Paradox}}, \href{https://doi.org/10.1080/001075199181387}{\emph{Contemp. Phys.} {\bfseries 40} (1999) 313}.

\bibitem{itzykson.zuber_1980_book_QuantumFieldTheory}
C.~Itzykson and J.B.~Zuber, \emph{Quantum {Field} {Theory}}, International {Series} in {Pure} and {Applied} {Physics}, McGraw-Hill, New York (1980).

\bibitem{bjorken.drell_1964_book_RelativisticQuantumMechanics}
J.D.~Bjorken and S.D.~Drell, \emph{Relativistic {Quantum} {Mechanics}}, International {Series} in {Pure} and {Applied} {Physics}, McGraw-Hill, New York (1964).

\bibitem{young.kim_2009_Nat.Phys._QuantumInterferenceKlein}
A.F.~Young and P.~Kim, \emph{Quantum {Interference} and {Klein} {Tunnelling} in {Graphene} {Heterojunctions}}, \href{https://doi.org/10.1038/nphys1198}{\emph{Nature Physics} {\bfseries 5} (2009) 222}.

\bibitem{stander.huard.ea_2009_Phys.Rev.Lett._EvidenceKleinTunneling}
N.~Stander, B.~Huard and D.~Goldhaber-Gordon, \emph{Evidence for {Klein} {Tunneling} in {Graphene} p-n {Junctions}}, \href{https://doi.org/10.1103/PhysRevLett.102.026807}{\emph{Phys. Rev. Lett.} {\bfseries 102} (2009) 026807}.

\bibitem{geim_2009_Science_GrapheneStatusProspects}
A.K.~Geim, \emph{Graphene: {Status} and {Prospects}}, \href{https://doi.org/10.1126/science.1158877}{\emph{Science} {\bfseries 324} (2009) 1530}.

\bibitem{geim.novoselov_2007_NatureMaterials_RiseGraphene}
A.K.~Geim and K.S.~Novoselov, \emph{The rise of graphene}, \href{https://doi.org/10.1038/nmat1849}{\emph{Nature Materials} {\bfseries 6} (2007) 183}.

\bibitem{castroneto.guinea.ea_2009_Rev.Mod.Phys._ElectronicPropertiesGraphene}
A.H.~Castro~Neto, F.~Guinea, N.M.R.~Peres, K.S.~Novoselov and A.K.~Geim, \emph{The electronic properties of graphene}, \href{https://doi.org/10.1103/RevModPhys.81.109}{\emph{Rev. Mod. Phys.} {\bfseries 81} (2009) 109}.

\bibitem{beenakker_2008_Rev.Mod.Phys._AndreevReflectionKlein}
C.W.J.~Beenakker, \emph{Andreev {Reflection} and {Klein} {Tunneling} in {Graphene}}, \href{https://doi.org/10.1103/RevModPhys.80.1337}{\emph{Rev. Mod. Phys.} {\bfseries 80} (2008) 1337}.

\bibitem{cooper.danjou.ea_2012_ISRNCond.Matt.Phys._ExperimentalReviewGraphene}
D.R.~Cooper, B.~D’Anjou, N.~Ghattamaneni, B.~Harack, M.~Hilke, A.~Horth et~al., \emph{Experimental {Review} of {Graphene}}, \href{https://doi.org/10.5402/2012/501686}{\emph{ISRN Cond. Matt. Phys.} {\bfseries 2012} (2012) 501686}.

\bibitem{abergel.apalkov.ea_2010_Adv.Phys._PropertiesGrapheneTheoretical}
D.~Abergel, V.~Apalkov, J.~Berashevich, K.~Ziegler,  and T.~Chakraborty, \emph{Properties of graphene: a theoretical perspective}, \href{https://doi.org/10.1080/00018732.2010.487978}{\emph{Adv. Phys.} {\bfseries 59} (2010) 261}.

\bibitem{schwinger_1951_Phys.Rev._GaugeInvarianceVacuum}
J.~Schwinger, \emph{On gauge invariance and vacuum polarization}, \href{https://doi.org/10.1103/PhysRev.82.664}{\emph{Phys. Rev.} {\bfseries 82} (1951) 664}.

\bibitem{fedotov.ilderton.ea_2023_Phys.Rep._AdvancesQEDIntense}
A.~Fedotov, A.~Ilderton, F.~Karbstein, B.~King, D.~Seipt, H.~Taya et~al., \emph{Advances in {QED} with intense background fields}, \href{https://doi.org/10.1016/j.physrep.2023.01.003}{\emph{Phys. Rep.} {\bfseries 1010} (2023) 1}.

\bibitem{ruffini.vereshchagin.ea_2010_Phys.Rep._ElectronPositronPairs}
R.~Ruffini, G.~Vereshchagin and S.-S.~Xue, \emph{Electron^^e2^^80^^93 {Positron} {Pairs} in {Physics} and {Astrophysics}: {From} {Heavy} {Nuclei} to {Black} {Holes}}, \href{https://doi.org/10.1016/j.physrep.2009.10.004}{\emph{Phys. Rep.} {\bfseries 487} (2010) 1}.

\bibitem{gelis.tanji_2016_Prog.Part.Nucl.Phys._SchwingerMechanismRevisited}
F.~Gelis and N.~Tanji, \emph{Schwinger {Mechanism} {Revisited}}, \href{https://doi.org/10.1016/j.ppnp.2015.11.001}{\emph{Prog. Part. Nucl. Phys.} {\bfseries 87} (2016) 1}.

\bibitem{schwartz_2013_book_QuantumFieldTheory}
M.D.~Schwartz, \emph{Quantum field theory and the standard model}, Cambridge University Press, Cambridge (2013).

\bibitem{greiner.muller.ea_1985_book_QuantumElectrodynamicsStrong}
W.~Greiner, B.~M^^c3^^bcller and J.~Rafelski, \emph{Quantum {Electrodynamics} of {Strong} {Fields}: {With} an {Introduction} into {Modern} {Relativistic} {Quantum} {Mechanics}}, Springer-Verlag, Berlin ; New York (1985).

\bibitem{fradkin.gitman.ea_1991_book_QuantumElectrodynamicsUnstable}
E.S.~Fradkin, D.M.~Gitman and S.M.~Shvartsman, \emph{Quantum {Electrodynamics} with {Unstable} {Vacuum}}, Springer {Series} in {Nuclear} and {Particle} {Physics}, Springer, Berlin Heidelberg (1991).

\bibitem{nikishov_1969_JETP_PairProductionConstant}
A.I.~Nikishov, \emph{Pair {Production} by a {Constant} {External} {Field}}, {\emph{JETP} {\bfseries 30} (1969) 660}.

\bibitem{nikishov_1970_Nucl.Phys.B_BarrierScatteringField}
A.~Nikishov, \emph{Barrier scattering in field theory: removal of {Klein} paradox}, \href{https://doi.org/10.1016/0550-3213(70)90527-4}{\emph{Nucl. Phys. B} {\bfseries 21} (1970) 346}.

\bibitem{nikishov_2004_Phys.Atom.Nucl._ScatteringPairProduction}
A.I.~Nikishov, \emph{Scattering and {Pair} {Production} by a {Potential} {Barrier}}, \href{https://doi.org/10.1134/1.1788038}{\emph{Phys. Atom. Nucl.} {\bfseries 67} (2004) 1478}.

\bibitem{hansen.ravndal_1981_Phys.Scr._KleinsParadoxIts}
A.~Hansen and F.~Ravndal, \emph{Klein's paradox and its resolution}, \href{https://doi.org/10.1088/0031-8949/23/6/002}{\emph{Phys. Scr.} {\bfseries 23} (1981) 1036}.

\bibitem{gavrilov.gitman_2016_Phys.Rev.D_ScatteringPairCreation}
S.P.~Gavrilov and D.M.~Gitman, \emph{Scattering and {Pair} {Creation} by a {Constant} {Electric} {Field} between {Two} {Capacitor} {Plates}}, \href{https://doi.org/10.1103/PhysRevD.93.045033}{\emph{Phys. Rev. D} {\bfseries 93} (2016) 045033}.

\bibitem{gavrilov.gitman_2016_Phys.Rev.D_QuantizationChargedFields}
S.P.~Gavrilov and D.M.~Gitman, \emph{Quantization of {Charged} {Fields} in the {Presence} of {Critical} {Potential} {Steps}}, \href{https://doi.org/10.1103/PhysRevD.93.045002}{\emph{Phys. Rev. D} {\bfseries 93} (2016) 045002}.

\bibitem{nakazato.ochiai_2022_Prog.Theor.Exp.Phys._UnstableVacuumFermion}
H.~Nakazato and M.~Ochiai, \emph{Unstable {Vacuum} and {Fermion} {Total} {Reflection} by the {Klein} {Step}}, \href{https://doi.org/10.1093/ptep/ptac085}{\emph{Prog. Theor. Exp. Phys.} {\bfseries 2022} (2022) 073B02}.

\bibitem{krekora.su.ea_2004_Phys.Rev.Lett._KleinParadoxSpatial}
P.~Krekora, Q.~Su and R.~Grobe, \emph{Klein {Paradox} in {Spatial} and {Temporal} {Resolution}}, \href{https://doi.org/10.1103/PhysRevLett.92.040406}{\emph{Phys. Rev. Lett.} {\bfseries 92} (2004) 040406}.

\bibitem{krekora.su.ea_2005_Phys.Rev.A_KleinParadoxSpinresolved}
P.~Krekora, Q.~Su and R.~Grobe, \emph{Klein paradox with spin-resolved electrons and positrons}, \href{https://doi.org/10.1103/PhysRevA.72.064103}{\emph{Phys. Rev. A} {\bfseries 72} (2005) 064103}.

\bibitem{cheng.su.ea_2010_Contemp.Phys._IntroductoryReviewQuantum}
T.~Cheng, Q.~Su and R.~Grobe, \emph{Introductory review on quantum field theory with space^^e2^^80^^93 time resolution}, \href{https://doi.org/10.1080/00107510903450559}{\emph{Contemp. Phys.} {\bfseries 51} (2010) 315}.

\bibitem{chervyakov.kleinert_2009_Phys.Rev.D_ExactPairProduction}
A.~Chervyakov and H.~Kleinert, \emph{Exact {Pair} {Production} {Rate} for a {Smooth} {Potential} {Step}}, \href{https://doi.org/10.1103/PhysRevD.80.065010}{\emph{Phys. Rev. D} {\bfseries 80} (2009) 065010}.

\bibitem{chervyakov.kleinert_2018_Phys.Part.Nuclei_ElectronPositronPair}
A.~Chervyakov and H.~Kleinert, \emph{On {Electron}^^e2^^80^^93{Positron} {Pair} {Production} by a {Spatially} {Inhomogeneous} {Electric} {Field}}, \href{https://doi.org/10.1134/S1063779618030036}{\emph{Phys. Part. Nuclei} {\bfseries 49} (2018) 374}.

\bibitem{schutzhold.gies.ea_2008_Phys.Rev.Lett._DynamicallyAssistedSchwinger}
R.~Sch^^c3^^bctzhold, H.~Gies and G.~Dunne, \emph{Dynamically {Assisted} {Schwinger} {Mechanism}}, \href{https://doi.org/10.1103/PhysRevLett.101.130404}{\emph{Phys. Rev. Lett.} {\bfseries 101} (2008) 130404}.

\bibitem{schneider.schutzhold_2016_J.HighEnerg.Phys._DynamicallyAssistedSauterSchwinger}
C.~Schneider and R.~Sch^^c3^^bctzhold, \emph{Dynamically {Assisted} {Sauter}-{Schwinger} {Effect} in {Inhomogeneous} {Electric} {Fields}}, \href{https://doi.org/10.1007/JHEP02(2016)164}{\emph{J. High Energ. Phys.} {\bfseries 2016} (2016) 164}.

\bibitem{torgrimsson.schneider.ea_2017_J.HighEnerg.Phys._DynamicallyAssistedSauterSchwinger}
G.~Torgrimsson, C.~Schneider, J.~Oertel and R.~Sch^^c3^^bctzhold, \emph{Dynamically {Assisted} {Sauter}-{Schwinger} {Effect} ― {Non}-{Perturbative} versus {Perturbative} {Aspects}}, \href{https://doi.org/10.1007/JHEP06(2017)043}{\emph{J. High Energ. Phys.} {\bfseries 2017} (2017) 43}.

\bibitem{torgrimsson.schneider.ea_2018_Phys.Rev.D_SauterSchwingerPairCreation}
G.~Torgrimsson, C.~Schneider and R.~Sch^^c3^^bctzhold, \emph{Sauter-{Schwinger} {Pair} {Creation} {Dynamically} {Assisted} by a {Plane} {Wave}}, \href{https://doi.org/10.1103/PhysRevD.97.096004}{\emph{Phys. Rev. D} {\bfseries 97} (2018) 096004}.

\bibitem{akal.egger.ea_2019_Phys.Rev.D_SimulatingDynamicallyAssisted}
I.~Akal, R.~Egger, C.~M^^c3^^bcller and S.~Villalba-Ch^^c3^^a1vez, \emph{Simulating {Dynamically} {Assisted} {Production} of {Dirac} {Pairs} in {Gapped} {Graphene} {Monolayers}}, \href{https://doi.org/10.1103/PhysRevD.99.016025}{\emph{Phys. Rev. D} {\bfseries 99} (2019) 016025}.

\bibitem{taya_2019_Phys.Rev.D_FranzKeldyshEffectStrongField}
H.~Taya, \emph{Franz-{Keldysh} {Effect} in {Strong}-{Field} {QED}}, \href{https://doi.org/10.1103/PhysRevD.99.056006}{\emph{Phys. Rev. D} {\bfseries 99} (2019) 056006}.

\bibitem{huang.taya_2019_Phys.Rev.D_SpinDependentDynamicallyAssisted}
X.-G.~Huang and H.~Taya, \emph{Spin-{Dependent} {Dynamically} {Assisted} {Schwinger} {Mechanism}}, \href{https://doi.org/10.1103/PhysRevD.100.016013}{\emph{Phys. Rev. D} {\bfseries 100} (2019) 016013}.

\bibitem{taya.fujimori.ea_2021_J.HighEnerg.Phys._ExactWKBAnalysis}
H.~Taya, T.~Fujimori, T.~Misumi, M.~Nitta and N.~Sakai, \emph{Exact {WKB} {Analysis} of the {Vacuum} {Pair} {Production} by {Time}-{Dependent} {Electric} {Fields}}, \href{https://doi.org/10.1007/JHEP03(2021)082}{\emph{J. High Energ. Phys.} {\bfseries 2021} (2021) 82}.

\bibitem{queisser.schutzhold_2019_Phys.Rev.C_DynamicallyAssistedNuclear}
F.~Queisser and R.~Sch^^c3^^bctzhold, \emph{Dynamically assisted nuclear fusion}, \href{https://doi.org/10.1103/PhysRevC.100.041601}{\emph{Phys. Rev. C} {\bfseries 100} (2019) 041601}.

\bibitem{kohlfurst.queisser.ea_2021_Phys.Rev.Res._DynamicallyAssistedTunneling}
C.~Kohlf^^c3^^bcrst, F.~Queisser and R.~Sch^^c3^^bctzhold, \emph{Dynamically assisted tunneling in the impulse regime}, \href{https://doi.org/10.1103/PhysRevResearch.3.033153}{\emph{Phys. Rev. Res.} {\bfseries 3} (2021) 033153}.

\bibitem{ryndyk.kohlfurst.ea_2024_Phys.Rev.Res._DynamicallyAssistedTunneling}
D.~Ryndyk, C.~Kohlf^^c3^^bcrst, F.~Queisser and R.~Sch^^c3^^bctzhold, \emph{Dynamically assisted tunneling in the {Floquet} picture}, \href{https://doi.org/10.1103/PhysRevResearch.6.023056}{\emph{Phys. Rev. Res.} {\bfseries 6} (2024) 023056}.

\bibitem{fradkin.gitman_1981_Fortschr.Phys._FurryPictureQuantum}
E.S.~Fradkin and D.M.~Gitman, \emph{Furry picture for quantum electrodynamics with pair-creating external field}, \href{https://doi.org/10.1002/prop.19810290902}{\emph{Fortschr. Phys.} {\bfseries 29} (1981) 381}.

\bibitem{sunakawa_1955_Prog.Theor.Phys._FormalTheoryScattering}
S.~Sunakawa, \emph{The formal theory of scattering}, \href{https://doi.org/10.1143/PTP.14.175}{\emph{Prog. Theor. Phys.} {\bfseries 14} (1955) 175}.

\bibitem{namiki.iino_1958_Prog.Theor.Phys.Suppl._NewMathematicalFormulation}
M.~Namiki and R.~Iino, \emph{A new mathematical formulation of quantum mechanics in the framework of wave-packet theory: {Transformation} theory, theory of scattering and functional differentiation}, \href{https://doi.org/10.1143/PTPS.5.65}{\emph{Prog. Theor. Phys. Suppl.} {\bfseries 5} (1958) 65}.

\bibitem{nitta.kudo.ea_1999_Am.J.Phys._MotionWavePacket}
H.~Nitta, T.~Kudo and H.~Minowa, \emph{Motion of a wave packet in the {Klein} paradox}, {\emph{Am. J. Phys.} {\bfseries 67} (1999) 966}.

\bibitem{leo.rotelli_2006_Phys.Rev.A_BarrierParadoxKlein}
S.D.~Leo and P.P.~Rotelli, \emph{Barrier paradox in the {Klein} zone}, {\emph{Phys. Rev. A} {\bfseries 73} (2006) 042107}.

\bibitem{ochiai.nakazato_2018_J.Phys.Commun._CompletenessScatteringStates}
M.~Ochiai and H.~Nakazato, \emph{Completeness of {Scattering} {States} of the {Dirac} {Hamiltonian} with a {Step} {Potential}}, \href{https://doi.org/10.1088/2399-6528/aa9fc0}{\emph{J. Phys. Commun.} {\bfseries 2} (2018) 015006}.

\bibitem{sauter_1932_Z.Phys._KleinschenParadoxon}
F.~Sauter, \emph{Zum ,,{Kleinschen} {Paradoxon}“}, \href{https://doi.org/10.1007/BF01349862}{\emph{Z. Phys.} {\bfseries 73} (1932) 547}.

\bibitem{katsnelson.novoselov.ea_2006_Nat.Phys._ChiralTunnellingKlein}
M.I.~Katsnelson, K.S.~Novoselov and A.K.~Geim, \emph{Chiral {Tunnelling} and the {Klein} {Paradox} in {Graphene}}, \href{https://doi.org/10.1038/nphys384}{\emph{Nature Physics} {\bfseries 2} (2006) 620}.

\bibitem{bai.zhang_2007_Phys.Rev.B_KleinParadoxResonant}
C.~Bai and X.~Zhang, \emph{Klein paradox and resonant tunneling in a graphene superlattice}, \href{https://doi.org/10.1103/PhysRevB.76.075430}{\emph{Phys. Rev. B} {\bfseries 76} (2007) 075430}.

\bibitem{pereira.peeters.ea_2010_Semicond.Sci.Technol._KleinTunnelingSingle}
J.M.~Pereira, F.M.~Peeters, A.~Chaves and G.A.~Farias, \emph{Klein tunneling in single and multiple barriers in graphene}, \href{https://doi.org/10.1088/0268-1242/25/3/033002}{\emph{Semicond. Sci. Technol.} {\bfseries 25} (2010) 033002}.

\bibitem{allain.fuchs_2011_Eur.Phys.J.B_KleinTunnelingGraphene}
P.E.~Allain and J.N.~Fuchs, \emph{Klein tunneling in graphene: optics with massless electrons}, \href{https://doi.org/10.1140/epjb/e2011-20351-3}{\emph{Eur. Phys. J. B} {\bfseries 83} (2011) 301}.

\bibitem{navarro-giraldo.quimbay_2020_Ann.Phys._TwoDimensionalKleinTunneling}
J.~Navarro-Giraldo and C.~Quimbay, \emph{Two-{Dimensional} {Klein} {Tunneling} for {Massive} {Dirac} {Fermions} with a {Defined} {Helicity}}, \href{https://doi.org/10.1016/j.aop.2019.168022}{\emph{Ann. Phys.} {\bfseries 412} (2020) 168022}.

\bibitem{schmitt.vallet.ea_2023_NaturePhysics_MesoscopicKleinSchwingerEffect}
A.~Schmitt, P.~Vallet, D.~Mele, M.~Rosticher, T.~Taniguchi, K.~Watanabe et~al., \emph{Mesoscopic {Klein}-{Schwinger} effect in graphene}, \href{https://doi.org/10.1038/s41567-023-01978-9}{\emph{Nature Physics} {\bfseries 19} (2023) 830}.

\bibitem{braun.su.ea_1999_Phys.Rev.A_NumericalApproachSolve}
J.W.~Braun, Q.~Su and R.~Grobe, \emph{Numerical approach to solve the time-dependent {Dirac} equation}, \href{https://doi.org/10.1103/PhysRevA.59.604}{\emph{Phys. Rev. A} {\bfseries 59} (1999) 604}.

\bibitem{mocken.keitel_2004_J.Comput.Phys._QuantumDynamicsRelativistic}
G.R.~Mocken and C.H.~Keitel, \emph{Quantum dynamics of relativistic electrons}, \href{https://doi.org/https://doi.org/10.1016/j.jcp.2004.02.020}{\emph{J. Comput. Phys.} {\bfseries 199} (2004) 558}.

\bibitem{mocken.keitel_2008_Comput.Phys.Commun._FFTsplitoperatorCodeSolving}
G.R.~Mocken and C.H.~Keitel, \emph{{FFT}-split-operator code for solving the {Dirac} equation in 2+1 dimensions}, \href{https://doi.org/https://doi.org/10.1016/j.cpc.2008.01.042}{\emph{Comput. Phys. Commun.} {\bfseries 178} (2008) 868}.

\end{thebibliography}\endgroup

\end{document}